%% file: data_paper_astroph.tex
\documentclass[12pt,preprint]{aastex}
\usepackage{graphicx}

\newcommand{\msunyr}{M$_{\odot}$~yr$^{-1}$}
\slugcomment{Re-submitted to ApJS 21 Sept 2010}

\shorttitle{GALEX UV Imaging of Local Volume Galaxies}
\shortauthors{Lee et al.}

\begin{document}

\title{A GALEX Ultraviolet Imaging Survey\\ of Galaxies in the Local Volume}

\author{
Janice C. Lee\altaffilmark{1}, 
Armando Gil de Paz\altaffilmark{2},
Robert C. Kennicutt, Jr.\altaffilmark{3,4},
Matthew Bothwell\altaffilmark{3},
Julianne Dalcanton\altaffilmark{5},
Jos\'e G. Funes, S.J.\altaffilmark{6},
Benjamin Johnson\altaffilmark{3},
Shoko Sakai\altaffilmark{7},
Evan Skillman\altaffilmark{8},
Christy Tremonti\altaffilmark{9},
Liese van Zee\altaffilmark{10}}

\altaffiltext{1}{Carnegie Starr Fellow, Carnegie Observatories, 813 Santa Barbara Street, Pasadena, CA 91101; jlee@obs.carnegiescience.edu}
\altaffiltext{2}{Departmento de Astrofisica, Universidad Complutense de Madrid, Madrid 28040, Spain}
\altaffiltext{3}{Institute of Astronomy, University of Cambridge, Madingley Road, Cambridge CB3 0HA, UK}
\altaffiltext{4}{Steward Observatory, University of Arizona, Tucson, AZ 85721}
\altaffiltext{5}{Department of Astronomy, University of Washington, Box 351580, Seattle, WA 98195}
\altaffiltext{6}{Vatican Observatory, Specola Vaticana, V00120 Vatican}
\altaffiltext{7}{Division of Astronomy and Astrophysics, University of California, Los Angeles, Los Angeles, CA, 90095-1562}
\altaffiltext{8}{Department of Astronomy, University of Minnesota, Minneapolis, MN 55455 }
\altaffiltext{9}{Department of Astronomy, University of Wisconsin-Madison, Madison, WI 53706}
\altaffiltext{10}{Astronomy Department, Indiana University, Bloomington, IN 47405}

\begin{abstract}

We present results from a GALEX ultraviolet (UV) survey of a complete sample of 390 galaxies within $\sim$11 Mpc of the Milky Way.  The UV data are a key component of the composite Local Volume Legacy (LVL), an ultraviolet-to-infrared imaging program designed to provide an inventory of dust and star formation in nearby spiral and irregular galaxies.  The ensemble dataset is an especially valuable resource for studying star formation in dwarf galaxies, which comprise over 80\% of the sample.  We describe the GALEX survey programs which obtained the data and provide a catalog of far-UV ($\sim1500$\AA) and near-UV ($\sim$2200\AA) integrated photometry.  General UV properties of the sample are briefly discussed.  We compute two measures of the global star formation efficiency, the SFR per unit HI gas mass and the SFR per unit stellar mass, to illustrate the significant differences that can arise in our understanding of dwarf galaxies when the FUV is used to measure the SFR instead of H$\alpha$.  We find that dwarf galaxies may not be as drastically inefficient in coverting gas into stars as suggested by prior H$\alpha$ studies.  In this context, we also examine the UV properties of late-type dwarf galaxies that appear to be devoid of star formation because they were not detected in previous H$\alpha$ narrowband observations.  Nearly all such galaxies in our sample are detected in the FUV, and have FUV SFRs that fall below the limit where the H$\alpha$ flux is robust to Poisson fluctuations in the formation of massive stars.  Otherwise, the UV colors and star formation efficiencies of H$\alpha$-undetected, UV-bright dwarf irregulars appear to be relatively unremarkable with respect to those exhibited by the general population of star-forming galaxies.

\end{abstract}


\section{Introduction}

Since the {\it Galaxy Evolution Explorer} (GALEX, Martin et al.$\;$2005) was successfully launched in 2003, the study of star formation in the ultraviolet (UV) 
has been broadly enabled and expanded to regimes previously unexplored.  
In particular, GALEX far-UV (FUV; $\sim1500$\AA) imaging has proved to be an 
exquisite probe of recent star formation in relatively dust-free galactic environments that are characterized by low density and low surface brightness.  

A major theme that has emerged from work based on GALEX data is that there may be more star formation in low density environments than previously recognized.  For example, an early surprise from the mission was the detection of FUV emission extending to several times the optical disks of the nearby systems M83 (Thilker et al. 2005) and NGC 4625 (Gil de Paz et al. 2005), and it was reported that the amount and extent of the star formation in these ``XUV-disks'' was greater than formerly seen in images of the H$\alpha$ nebular emission-line.  Subsequent studies, which have begun to constrain the prevalence of XUV-disks in the overall population, have indicated that extended star formation is common, and occurs in roughly one-quarter to one-third of disk galaxies (Thilker et al. 2007; Zaritsky \& Christlein, 2007; Goddard et al. 2010). 

Later work investigated the globally integrated UV star formation rates (SFR) of local dwarf and low surface brightness galaxies, and several groups showed that, again, the amount of activity inferred from the UV in such systems tends to be greater than would be expected from previous H$\alpha$ measurements (e.g., Meurer et al. 2009; Lee et al. 2009b; Boselli et al. 2009; Hunter et al. 2010).  Our own study --- based on the UV dataset presented in this paper --- was the first to establish this systematic trend with a statistically significant sample of dwarf galaxies with SFRs from 0.1 \msunyr~ down to 0.0001 \msunyr~ (Lee et al. 2009b).  We find that at SFR$\sim$0.003 \msunyr, the average H$\alpha$-to-FUV flux ratio is lower than expected by about a factor of two, and at the lowest SFRs probed by our sample, the ratio exhibits up to an order of magnitude discrepancy.  Prior to our work, there was a lack of deep UV observations for a complete, well-defined sample of dwarf galaxies, and this had prevented the trend from being clearly delineated in the past. 

Until GALEX, a great deal of our knowledge of star formation in local galaxies had been based on H$\alpha$ observations (e.g., Hodge 1974; Kennicutt 1989; Kennicutt et al. 2008 and references therein).  From the recent GALEX results just summarized however, it is now thought that the UV may be a more robust diagnostic of star formation activity in the low SFR intensity regime.  UV emission should be less prone to stochastic effects from sparse sampling of the upper end of the stellar initial mass function (IMF), and to possible uncertainties in the fate of ionizing photons in low-density gas; that is, whether the HII regions, or even entire galaxies, can be safely assumed to be ionization-bounded (e.g., Melena et al. 2009).  This is because the UV flux is directly emitted from the photospheres of O- through later-type B-stars ($M_{\ast}\gtrsim3 M_{\odot}$), whereas the H$\alpha$ nebular emission arises from the recombination of hydrogen which can only be ionized by the most massive O- and early-type B-stars ($M_{\ast}\gtrsim17 M_{\odot}$).  However, it is currently debated whether stochasticity or the potential leakage of Lyman continuum photons can fully account for the magnitude of the observed systematic between UV and H$\alpha$ SFRs.  A range of other causes have been considered (e.g., dust attenuation, metallicity, discontinuous star formation histories),  the most debated of which is an IMF deficient in the most massive stars (Pflamm-Altenburg et al. 2009 and references therein; Meurer et al. 2009; Lee et al. 2009b; Boselli et al. 2009; Hunter et al. 2010).  Though the explanation for the discrepancies between the two tracers is still uncertain, it is clear that UV observations from GALEX have  
challenged our understanding of star formation in low density environments, and are {\it essential} for the study of the outer disks of spirals, and of low-surface brightness and dwarf galaxies.   

In this paper, we present results from a GALEX survey of a complete sample of Local Volume galaxies whose primary defining characteristic is that it is dominated by dwarf galaxies --- over 80\% of the sample have luminosities and star formation rates lower than those of the Large Magellanic Cloud.   This effort was undertaken as part of broader multi-wavelength campaign to provide new insight into the mechanisms that drive, regulate and extinguish global star formation in galaxies, with special focus on the dwarf and low surface brightness regime.  Our survey has sought to build upon the patchwork of existing observations of nearby galaxies, which suffer from common observational biases toward luminous, high surface brightness systems (e.g., Blanton et al. 2005). 
Though a great deal has already been learned about the nature of nearby dwarf star-forming galaxies, as low-metallicity, gas-rich, but not necessarily young systems 
(e.g.; Gallagher \& Hunter 1984; van Zee 2000; Gil de Paz et al. 2003; Hunter \& Elmegreen 2004), the vast majority of prior studies have relied on observations of representative samples of dwarfs.  Our work seeks to enable studies which demand dwarf galaxy samples that are not only representative,
but are true to the statistics provided by an approximately volume-limited sample (e.g., characterization of the prevalence of starbursts in low mass systems; Lee et al. 2009a).

H$\alpha$ narrowband imaging observations for the sample were previously carried out by Kennicutt et al. (2008).  The precursor H$\alpha$ survey and GALEX follow-up program together comprise 11HUGS, the 11 Mpc H$\alpha$ and Ultraviolet Galaxy Survey. 
The data from 11HUGS have further been augmented by Spitzer IRAC mid-infrared and MIPS far-infrared observations (Dale et al. 2009) through the composite Local Volume Legacy\footnote{http://www.ast.cam.ac.uk/research/lvls/} (LVL) program, which has provided the ensemble UV, H$\alpha$ and IR dataset to the community through the NASA/IPAC Infrared Science Archive\footnote{http://ssc.spitzer.caltech.edu/spitzermission/observingprograms/legacy/lvl/} (IRSA).  11HUGS largely focuses on star-forming systems brighter than a certain limit ($B<15.5$).  To extend the coverage of the galaxy population to dwarf spheroidals, dwarf ellipticals and the faintest irregular galaxies, such systems from the ACS Nearby Galaxy Survey Treasury (ANGST) program (Dalcanton et al. 2009) were also targeted for both GALEX and Spitzer observations.  ANGST has obtained Hubble Space Telescope resolved stellar population imaging for a volume-limited sample outside the Local Group and within $\sim$4 Mpc.  Altogether, the GALEX dataset presented in this paper provides the most complete catalog of integrated UV photometry for Local Volume galaxies currently available.

The remainder of the paper is organized as follows.
In Section 2, we describe in detail the resultant Local Volume sample selected for GALEX follow-up, summarize the observations, and present a catalog of NUV and FUV integrated photometry.  In Section 3, basic results from the survey are presented.  We give an overview of the UV properties of the sample and discuss the detection rate.  We compute two measures of the global star formation efficiency, the SFR per unit HI gas mass and the SFR per unit stellar mass, to illustrate the significant differences that can arise in our understanding of dwarf galaxies when the FUV is used to measure the SFR instead of H$\alpha$.  In particular, the UV properties of late-type dwarf galaxies that appear to be devoid of star formation because they were not detected in previous H$\alpha$ narrowband observations are examined.  We comment on the selection of ``transition'' dwarf galaxies (dIrr/dSph) based upon the non-detection of H$\alpha$ emission in gas-rich dwarfs.  We adopt $H_o=75$ km s$^{-1}$ Mpc$^{-1}$ for distance-dependent quantities, when distance measurements from standard candles or secondary indicators are not available.

\section{Data}

\subsection{Sample Selection}

The Local Volume galaxies that were targeted for GALEX imaging were mainly selected from the sample given in Kennicutt et al. (2008; hereafter Paper I), who have carried out a deep, statistically complete, H$\alpha$+[NII] narrowband imaging survey within $\sim$11 Mpc.  The ANGST program (Dalcanton et al. 2009), which has obtained HST ACS and WFPC2 imaging for a $\sim$4 Mpc volume-limited sample, contains about 20 low-luminosity and/or early-type galaxies that were not already included in Paper I, and GALEX data were also obtained for these objects.  In this section, we first summarize the Paper I sample, and then describe the resultant set of galaxies that were observed by GALEX in our survey. 

Paper I presents a sample of 436 galaxies from existing catalogs which were compiled in two components.  The primary component aims to be as complete as possible in its inclusion of known nearby star-forming galaxies within given limits.  It consists of all known spirals and irregulars (T$\geq$0) within 11 Mpc that avoid the Galactic plane ($|b|>20^{\circ}$) and are brighter than $B=15$ mag.  These particular limits represent the ranges within which the original surveys that have provided the bulk of our knowledge of the Local Volume galaxy population have been shown to be relatively complete (e.g., Tully 1988, deVaucouleurs et al. 1991), but still span a large enough volume to probe a diverse cross-section of star formation properties.  A secondary component consists of galaxies that are within 11 Mpc for which H$\alpha$ flux measurements are available, but fall outside one of the limits on brightness, Galactic latitude and morphological type --- these were generally targets either observed by our group as telescope time allowed, or had H$\alpha$ fluxes published in the literature.  Subsequent statistical tests, as functions of the compiled B-band apparent magnitudes and 21-cm (HI) fluxes, show that the overall sample is complete to $\sim$15.5 mag and $\sim$6 Jy km s$^{-1}$, respectively.  This corresponds to limits of $M_B\lesssim-15$ and $M_{HI}\gtrsim 2 \times 10^8$ M$_{\odot}$ for $|b|>20^{\circ}$ at the edge of the 11 Mpc volume.  More details on the parent sample and its properties are given in Paper I (sample construction, Local Volume membership uncertainties, and integrated H$\alpha$ flux and equivalent width catalog), Lee et al. (2007; overall star formation demographics as traced by the H$\alpha$ equivalent width), and Lee et al. (2009a; completeness tests and limits, and dwarf galaxy starburst statistics).

Galaxies with $|b|>30^{\circ}$ and $B\leq15.5$ from the Paper I sample (N=256) were targeted for UV imaging through the 11HUGS GALEX Legacy program (GI1\_047, GI4\_095).  The Galactic latitude limit was imposed to avoid excessive foreground extinction, and fields with bright stars and/or high background levels for which imaging would be prohibited due to the instrument's brightness safety limits.  GALEX observations for a significant number (N=120) of the remaining lower latitude, fainter galaxies in the parent sample have also been taken by other programs and are publicly available in the GALEX archive through MAST.\footnote{MAST is the Multimission Archive at the Space Telescope Science Institute (http://archive.stsci.edu/index.html).  STScI is operated by the Association of Universities for Research in Astronomy, Inc., under NASA contract NAS5-26555. Support for MAST for non-HST data is provided by the NASA Office of Space Science via grant NAG5-7584 and by other grants and contracts.}  We have performed photometry on these data as well, and provide measurements in the tables that follow.

Observations of the 11HUGS galaxies have also been extended into the infrared using the Spitzer Space Telescope as part of the Local Volume Legacy (LVL) survey (Dale et al. 2009).  As mentioned above, in addition to galaxies selected from Paper I, LVL also includes targets from ANGST in order to extend coverage of galaxy properties to dwarf spheroidal and ellipticals, and the faintest dwarf irregulars known outside of the Local Group.  There are 27 galaxies in the composite LVL sample which are not included in Paper I.\footnote{\label{footnote}Included in this group of 27 are 4 galaxies for which a revision in the distance estimates placed them outside 11 Mpc after the final sample for LVL Spitzer observations had been selected (see section 2.2).}  
GALEX observations have also been obtained for all of these galaxies, about half of which were taken through a GALEX follow-up program of ANGST galaxies (GI3\_06).  The photometry for these 27 galaxies is also reported here.  

\subsection{GALEX Observations}

GALEX is a NASA Small Explorer launched in 2003 April.  It uses a 50 cm aperture telescope with a dichroic beam splitter that enables simultaneous observations in the FUV ($\lambda_{eff}$=1539\AA, FWHM=269\AA) and the NUV ($\lambda_{eff}$=2316\AA, FWHM=616\AA).  The field-of-view of the camera is circular and has a 1\fdg2 diameter.  The PSF is dependent on the count rate (deteriorates for brighter sources) as well as on the radial position on the image (deteriorates at the field edge), but is reasonably well described by a FWHM of 5$\pm$1\arcsec\ for the data analyzed here.  The images that we use have been uniformly reprocessed by the GALEX team with the most recent version of the pipeline (v6) available at the time of the initial preparation of this paper.  The v6 pipeline accounts for a drift in the photometric calibration (1.5\% and 0.25\% dimmer per year for the NUV and FUV bands, respectively), incorporates corrections to the flat-fielding (5\% systematic in the FUV and a 2\% change in shape in the NUV), and improves removal of reflected light at the NUV field edge.  Full details on the satellite, telescope, instrument and calibration and data processing pipeline are given in Martin et al. (2005), Morrissey et al. (2005), Morrissey et al. (2007).  
 
Our primary survey objective is to provide a statistically complete ultraviolet dataset for a deep Local Volume galaxy sample.  To do this, our program leverages upon existing data, and homogeneously fills significant gaps in prior GALEX coverage of the $D<11$ Mpc, $b>30^{\circ}$, $B<15.5$ population (N=256).  The GALEX team's Nearby Galaxy Survey and Medium Imaging Survey has supplied imaging for $\sim$30\% of this sample (Gil de Paz et al. 2007), with cumulative exposure times of $\sim$1500 sec per field (close to the time available in one orbit), while a small fraction ($\sim$10\%) were observed to similar depth by other independent guest investigator programs.  This exposure time corresponds to surface brightness limits of $m_{AB}\sim27.5$ mag arcsec$^{-2}$ or an SFR of $\sim$10$^{-3}$ \msunyr$\;$kpc$^{-2}$ (Martin et al. 2005; Gil de Paz et al. 2007).  Our Cycle 1 (GI1-047) and Cycle 4 (GI4-095) programs obtained one-orbit depth data for an additional $\sim$50\% of the sample.  The status of GALEX observations for the remaining of $\sim$10\% of this sample is as follows.  Observations for 10 galaxies are prohibited due to bright foreground stars and/or high background (LMC, SMC, ISZ399, NGC1313, NGC3077, NGC7713, UGC5076, UGC7490, UGC8508, UGC8837).\footnote{Although observations for fields close to NGC1313 and NGC3077 are currently not allowed, some early data were taken by the GALEX team for these galaxies.  A shallow All-sky Imaging Survey (AIS) 211 sec pointing exists for NGC1313.  Observations of NGC3077 were attempted by the Nearby Galaxy Survey, but in the effort to shift the offending bright stars off of the GALEX FOV, the galaxy was moved too close to the field edge, and only about half of it was captured in the resultant imaging.}  Observations for 10 galaxies have not been completed to sufficient depth in the FUV (i.e., less than 1/3 the requested integration time, or 500 sec), and only shallow imaging from the GALEX All-sky Imaging Survey (AIS) are available (IC625, IC2782, IC4316, NGC3675, NGC5608, UGC5451, UGC8091, UGC9128, UGC9211, UGC9660).\footnote{Two of these galaxies, IC2782 and UGC9128 have deep ($>$1500 sec) exposures in the NUV, though their FUV data are shallow.  This is a consequence of intermittent FUV detector failures, during which observations still proceeded in the NUV channel.}  No data are available for 4 galaxies (ESO252-IG001, ESO306-G013, UGC5151, UGCA103).  Overall, observations for the ensemble dataset were taken over five years between 2004-2009.  Intermittent FUV detector failures, or ``transient overcurrent events,'' that have occurred since 2004 (Morrissey 2006), caused the completion of the survey to require more time than initially anticipated.

UV imaging is also available from other GALEX programs for another 120 of the remaining 180 galaxies in Paper I which fall outside the limits of the main $b>30^{\circ}$ and $B\leq15.5$ sample.  Approximately half of these galaxies have shallow AIS imaging ($\lesssim$200 sec), while the remainder have deep ($\geq$1500 sec) exposures. 

Figure 1 summarizes the resultant GALEX coverage of the overall 11 Mpc sample as described above.  Distributions of the morphological type, $B$ magnitude, Galactic latitude
and distance (i.e., the properties used to define our samples) are shown.  
The overall 11 Mpc parent sample (Table 1) is given by the gray histogram, while the sub-set selected for GALEX observations by 11HUGS ($b>30^{\circ}$, $B\leq15.5$) is distinguished in white.  The hatched portions of the histograms show the fractions of each sample for which only shallow imaging exists ($t_{FUV}<200$ sec), while the black portions show those for which no FUV observations have been carried out.  There are no strong biases in the available one-orbit depth GALEX imaging in any of these properties.

Table 1 lists specific exposure times and information on the original GALEX programs that obtained the data, along with general properties of the targets.  To facilitate cross-comparison with the H$\alpha$ data presented in Paper I, basic properties given in the first table of Paper I are repeated in columns 1--9 with some updates, and all 436 galaxies are listed, whether they have been observed by GALEX or not.  Galaxies in the composite LVL sample which are not included in Paper I (generally targets drawn from the ANGST sample) are listed separately at the end of the table.  The columns in Table 1 are as follows: 

\noindent Column (1): The running index number.

\noindent Column (2): Galaxy name.

\noindent Columns (3-4): J2000 Right Ascension and Declination as reported in NED.

\noindent Column (5): Galactic latitude as reported in NED.

\noindent Column (6): Adopted distance.  Measurements based on standard candles or secondary distance indicators were compiled from the literature when available, otherwise distances computed from recessional velocities corrected to the centroid of the Local Group (following Karachentsev \& Makarov 1996, as reported on NED) and adopting H$_{\circ}$= 75 km s$^{-1}$ Mpc$^{-1}$ are listed.  Original references for the direct distances are given in Paper I, Table 1, with the exception of galaxies targeted by the ANGST program, for which non-Cepheid distances have been updated to be consistent with the tip of the red giant branch measurements reported in Dalcanton et al. (2009).

\noindent Column (7): Distance determination method, with the following abbreviations: Cepheid variables (ceph), tip of the red giant branch (trgb), surface brightness fluctuations (sbf), membership in a group with measured distance (mem), brightest blue stars (bs), Tully-Fisher relation (tf), and Local Group corrected recessional velocity distances (flow).

\noindent Column (8): Apparent $B$-band magnitude.  The photometry were compiled as discussed in Lee et al. (2009a), and references are given in Paper I, Table 1.

\noindent Column (9): RC3 morphological type.

\noindent Column (10):  A flag indicating whether the galaxy is included in the {\it Spitzer} Local Volume Legacy program (1), or not (0).  The same $|b|>30$\degr, $B<15.5$ Local Volume population was targeted for both GALEX and {\it Spitzer}  (IRAC 3.6, 4.5, 5.8, and 8.0 $\mu$m and MIPS 24, 70, and 160 $\mu$m) follow-up.  However, changes in the adopted distances for some galaxies which occurred between the selection of the {\it Spitzer} and final GALEX/Paper I samples led to slight differences between the samples.  This flag is provided here to help clarify these differences.  As described in Dale et al. (2009), four galaxies targeted for {\it Spitzer} observations have updated distances which place them outside of 11 Mpc.  These galaxies appear at the end of the table.  The flow model initially applied was also updated to provide consistency with one used by the NASA/IPAC Extragalactic Database (NED).  As a result, 30 Paper I galaxies with $|b|>30$\degr, $B<15.5$ do not have {\it Spitzer} imaging in Dale et al. (2009). The galaxies are generally between 10 and 11 Mpc, where flow distance uncertainties ($\pm$15\%) may scatter objects in and out of the volume. Such uncertainties, however, should not be significant for studies seeking to use the sample to statistically characterize the physical properties of local galaxies. As we previously commented in Paper I and in Dale et al. (2009), an inherent difficulty with efforts to construct a local volume-limited sample is that its membership will necessarily be fluid until accurate distance and photometric measurements are available for all of the galaxies that are within the volume and around its periphery.  

\noindent Both the archival and newly obtained {\it Spitzer} imaging for the LVL sample have been uniformly processed (Dale et al. 2009).  The full set of IRAC, MIPS and available H$\alpha$ and GALEX UV imaging for the LVL galaxies are all provided at: 

http://ssc.spitzer.caltech.edu/spitzermission/observingprograms/legacy/lvl/ 

\noindent UV photometry measured in apertures identical to those used for the {\it Spitzer} photometry and spectral energy distributions published in Dale et al. (2009) is provided in Table 3.

\noindent Column (13): GALEX ``tile'' name(s) giving information on the original program that requested observations for the galaxy.

\subsection{GALEX Photometry}

The procedures used to perform photometry closely follow those used by Gil de Paz \& Madore (2005), for ground-based optical imaging, and Gil de Paz et al. (2007) for the GALEX Atlas of Nearby Galaxies.  Both total (asymptotic) magnitudes based on curve-of-growth analyses and aperture magnitudes are measured and reported in Tables 2 and 3.  Details on the contents of the tables are given at the end of this section.

To prepare the images for photometry, foreground stars and background galaxies are masked in a two-step procedure.  The IRAF task {\sc starfind} is used to identify all point sources down to 4$\sigma$ significance (i.e., at the GALEX resolution, $\sim$5\arcsec), and those with FUV$-$NUV$>$1 are flagged as foreground Galactic stars.  These initial masks are then inspected by eye.  The nuclei of a few early-type dwarf galaxies that were flagged by {\sc starfind} are unmasked.  Blue foreground stars are also identified during this inspection and added to the mask.  We note that such stars are rare at the Galactic latitudes targeted by 11HUGS ($|b|>30^{\circ}$), and are readily differentiated from stellar clusters or associations in the target galaxy.
Background galaxies were also identified by visual inspection, based upon the morphology in available optical imaging and/or anomalous UV color.  The masked pixels are replaced by values interpolated from the surrounding pixels.

To perform surface photometry and obtain total (asymptotic) magnitudes, knowledge of the sky background level is required.  We measure the background  
using pixels in equal-area
regions distributed in azimuth around the galaxy in two concentric
annuli, and compute the average sky 
background, local background standard deviation, and standard deviation 
of the mean across the field. 
The annuli are chosen to lie well beyond the RC3 (Third Reference Catalog of Bright Galaxies; de Vaucouleurs et al. 1991) $B$-band 25 mag 
arcsec$^{-2}$ isophote (typically
2-4 times $D_{25}$), where no emission from the
galaxy is detected.  The image used in this calculation is completely
masked of all objects, including the target galaxy itself.  The masks
for this sky calculation are created from objects detected in the GALEX pipeline and 
extracted from the GALEX merged catalog file, {\sc mcat}.

The surface photometry is carried out on the masked, locally interpolated 
image using the IRAF task {\sc ellipse},
with fixed central coordinates, ellipticity, and position angle.
The central coordinates are given in Table 1.  The position angles, 
and the semi-major and semi-minor axis sizes on which the ellipticities are based,
are given in Table 2.
Values for these parameters are drawn from 
NED and are generally based on $B$-band measurements published in the RC3.
The resulting elliptical apertures were individually checked by eye to 
ensure that they adequately described the galaxy on optical DSS images, 
and were correctly centered on the GALEX images.  Manual adjustments 
to the values taken from NED were made as necessary.

Surface brightness profiles are computed by {\sc ellipse}, where
errors in the azimuthally-averaged surface brightnesses
are taken into account, and include contributions from the photon noise and 
both high- and low-frequency errors in the
background error budget (see Gil de Paz et al$.$ 2007 and Gil de Paz
\& Madore 2005 for more details).  To determine the asymptotic
magnitudes, the growth curve in each GALEX UV band is computed.
A linear fit to the cumulative
magnitude versus the cumulative-magnitude gradient is performed, and the
y-intercept of the fit is adopted as the asymptotic magnitude. 
The uncertainty in the fit is also included in the reported errors.
This procedure is relatively standard at optical wavelengths 
(see e.g., Cair\'{o}s et al$.$ 2001), and has been shown
to be accurate for the determination of total magnitudes in 
GALEX UV imaging of nearby galaxies (Gil de Paz et al$.$ 2007).

Aperture fluxes are measured within the outermost elliptical 
annulus where both FUV and NUV surface photometry can be performed.
We define this annulus as the one beyond which either 
the flux error becomes larger than 0.8 mag
or where the intensity falls below that of the sky
background in both FUV and NUV bands.    These
matched aperture fluxes are appropriate for computing the integrated color.  

Finally, fluxes are also measured in the apertures used for
the LVL Spitzer imaging in Dale et al. (2009), to provide 
a complete set of matched UV-to-IR aperture photometry
as mentioned above.

\noindent The results of our photometry, on the AB magnitude scale,
are presented in Table 2 as follows.

\noindent Column (1): Index number, repeated from Table 1.

\noindent Column (2): Galaxy name, repeated from Table 1.

\noindent Column (3): E(B-V) based on the maps of Schlegel et al. (1998).

\noindent Column (4-6): Semi-major axis, semi-minor axis and position angle of the outermost elliptical annulus where {\it both} FUV and NUV surface photometry can be performed.  We define this as the distance beyond which the annular flux error becomes larger than 0.8 mag, or where
 the intensity falls below that of the sky background in both FUV and NUV bands, whichever occurs first.  The position angle is measured east of north as usual.

\noindent Column (7-10): FUV and NUV magnitude and error, measured in the common elliptical aperture described by the preceding columns.
Corrections for Milky Way foreground extinction have been applied using the $E(B-V)$ values listed here, and the Cardelli et al. (1989) extinction law with $R_V=3.1$.  That is, $A_{FUV}=7.9E(B-V)$ and $A_{FUV}=8.0E(B-V)$.   For non-detections, 3$\sigma$ point source upper-limits are listed, enclosed in parenthesis.  The quoted errors include the photometric uncertainty within the aperture 
and the error in the zeropoint (0.05 and 0.03 $m_{AB}$ in the FUV and NUV, respectively; Morrissey et al. 2007).

\noindent Column (11-12): FUV-NUV color and error, corrected for MW extinction as described previously.  

\noindent Column (13-16): FUV and NUV asymptotic magnitude and error, measured from a growth curve which is computed using fixed central coordinates, ellipticity and position angle as described above, and corrected for MW extinction.  
Here, the quoted errors include the photometric uncertainty, 
the error in the zeropoint, as well as the 
the uncertainty due to the error-weighted fit of the growth curve.

\noindent In Table 3, we list only those galaxies in the LVL sample, providing:

\noindent Column (1): Index number, as in Table 1.

\noindent Column (2): Galaxy name, as in Table 1.

\noindent Column (3-5): Semi-major axis, semi-minor axis and position angle of the elliptical annulus 
used by Dale et al. (2009) to extract photometry from Spitzer IRAC and MIPS imaging, reproduced
from Dale et al. (2009) Table 1.

\noindent Column (6-9): FUV and NUV magnitude and error, measured in the common elliptical aperture described by the preceding columns.
Corrections for Milky Way foreground extinction have been applied using the $E(B-V)$ values listed here, and the Cardelli et al. (1989) extinction law with $R_V=3.1$.  That is, $A_{FUV}=7.9E(B-V)$ and $A_{FUV}=8.0E(B-V)$.   For non-detections, 3$\sigma$ point source upper-limits are listed, enclosed in parenthesis.  The quoted errors include the photometric uncertainty within the aperture and the error in the zeropoint (0.05 and 0.03 $m_{AB}$ in the FUV and NUV, respectively; Morrissey et al. 2007).

\section{Results \& Discussion}

\subsection{Basic Sample Properties}\label{section:sample_properties}

The GALEX observations presented here significantly improve the depth to which complete, statistical sampling of Local Volume star-forming (a.k.a., ``blue sequence'') galaxies is achieved.  This is illustrated in Figure~\ref{fig:uvlf}, which shows the number distributions of FUV and NUV absolute magnitudes.  The filled histograms represent GALEX's resultant coverage of the primary 11HUGS target volume ($|b|<30$\degr, $d<11$ Mpc) prior to our observations, and the solid line shows the distributions when our observations are included.  While the more limited sample with prior coverage begins to become incomplete for $\gtrsim$-15 mag, our sample continues to increase until about $-$13 mag.  11HUGS mitigates the previous $\sim$4-fold under-representation of Local Volume galaxies at low luminosities.

Luminosity functions based on surveys that have well-determined selection functions and cover much larger volumes than 11HUGS also provide an interesting point of comparison for our $M_{FUV}$ and $M_{NUV}$ number distributions.  
Such luminosity functions are subject to typical biases toward the most luminous, massive, high surface brightness systems, and therefore are generally well-sampled at the bright-end, but are not robustly constrained in the dwarf galaxy regime.  In Figure~\ref{fig:uvlf} we overplot UV luminosity functions,\footnote{The luminosity functions were rescaled to reflect H$_{\circ}$= 75 km s$^{-1}$ Mpc$^{-1}$ as is assumed here to calculate distances for objects without standard candle/ruler estimates.} as computed by Schiminovich et al. (2007), for disk galaxies in the SDSS DR4 spectroscopic sample (Adelman-McCarthy et al. 2006) that have also been observed in the GALEX Medium Imaging Survey (Martin et al. 2005).  The curves give their best fitting Schechter function fits, with the dashed segments representing extrapolations past the last data point shown.  To force approximate agreement between the number distributions and the Schiminovich et al. (2007) luminosity functions at $\sim L^{*}$, the latter has been multiplied by 2 times the $|b|>30$, $d<11$ Mpc 11HUGS target volume.  The factor of 2 implies that the portion of the Local Volume probed by 11HUGS is over-dense compared with cosmological volumes, as discussed in Karachentsev et al. (2004) and Lee et al. (2009a).  Qualitatively, there is good agreement between the relative shapes of both $M_{FUV}$ and $M_{NUV}$ luminosity functions and the 11HUGS number distributions prior to the sharp fall-off of galaxies fainter than $\sim-$13 mag.  This provides some assurance that the 11HUGS sampling of bright galaxies, while sparse due to the small target volume, is not heavily biased relative to the global distribution of UV luminosities.  The extrapolated faint-end slopes of the luminosity functions appear to also be reasonable in comparison with our local dataset.  In $M_{FUV}$, the 11HUGS dataset appears to prefer a slightly steeper rise at the faint end ($\alpha\sim$1.06; gray), though this is well within the uncertainties of the value of $\alpha$=1.02$\pm$0.05 reported by Schiminovich et al. (2007).  If the standard SFR conversion recipe of Kennicutt (1998) is applied to the FUV luminosities, the SFRs probed by the sample span roughly 6 orders of magnitude, from $\sim$10 \msunyr to $\sim$10$^{-6}$ \msunyr. 

The dominance of low-luminosity star-forming galaxies in the sample is also illustrated by distributions in the $FUV$-$NUV$ color (corrected for Milky Way reddening, but not for dust attenuation internal to the galaxies themselves).  In Figure~\ref{fig:colormag}, we show a histogram of the UV color, and also plot it as a function of EW(H$\alpha$), $M_{FUV}$ and $M_B$.  The points are coded to indicate different morphological types, and the histogram is shaded in a corresponding manner.  The sample is concentrated in a fairly narrow range of UV color: the median is 0.29 mag, and 79\% of the galaxies have 0$<FUV$-$NUV<$0.5.  Such blue colors are primarily a result of the recent star formation activity in such systems (e.g., van Zee 2001; Hunter \& Elmegreen 2004; Lee et al. 2007), while low metallicities (e.g., Skillman et al. 1989; van Zee et al. 1997; Lee et al. 2004) and low dust attenuation (e.g., Lee et al. 2009b) also play a role.

Late-type spiral and irregular ``blue sequence'' galaxies can be roughly separated from early-type ``red sequence'' galaxies
using a division at $FUV$-$NUV$=0.9 (plotted as a dashed line in Figure~\ref{fig:colormag}) as described in Gil de Paz et al. (2007). 
Again, by design our sample does not contain many early-type galaxies, and 
there are few galaxies with $FUV$-$NUV>0.9$. 
For the star-forming galaxies with $FUV$-$NUV<$0.9, the UV color does not show strong trends with either $M_{FUV}$ or $M_B$, except perhaps for slightly redder values at $M_B$ brighter than -19, at which the sample is predominantly composed of earlier type spirals. 
There is also a weak trend towards redder colors at lower equivalent width.  This trend is expected, since EW(H$\alpha$) traces the current SFR relative to the past lifetime average, which provides a measure of the recent star formation history.

\subsection{GALEX Non-Detections}

Late-type galaxies in the Local Volume that do not have measurable 
UV emission appear to very rare, at least among the $B<15.5$ population 
where our observations are concentrated (see Figure 1).  
Only 22 of the 390 galaxies in our sample which were observed by
GALEX were not detected in the FUV.  About half of these (N=12) 
are galaxies classified as faint dwarf spheroidals/ellipticals, 
and would not be expected to have UV emission from recent star
formation.\footnote{Sculptor-dE1, UGCA133, FM2000-1, LEDA166101, KKH057, BK05N, UGC5442, IKN, DDO78, BK06N, [KK98]208, KKR25}  
Another 8 are classified as late-type dwarfs, 
but only have shallow imaging ($\lesssim$200 s) available; 
in these cases a detection is also not expected given the 
galaxies' low $B$-band luminosities and/or low H$\alpha$-based
SFRs.\footnote{IC2782, LSBCD565-06, LSBCD634-03, UGCA86, KKH34, UGCA92, HS[98]117, LEDA166115}
The remaining two, UGC5428 and UGCA276 do have deep GALEX imaging (t$\sim$1700 s), 
and are cases for which further follow-up may be valuable as described further below.  
Both have extremely low luminosities with $M_B\sim-12$, and are nearby (D$\sim$3 Mpc).
The GALEX imaging of these galaxies is sensitive enough to detect a single early-type B star
($\sim$15 M$_{\odot}$).

UGC5428 is a member of the M81 group, and has been classified as a ``transition'' 
dwarf -- a system whose properties are intermediate between those of late 
and early-type galaxies (e.g., Mateo 1998; Grebel et al. 2003).
Although it is seen to be dominated by an old stellar population 
based upon ANGST HST imaging (Weisz et al. 2010), 
and HI gas has not been observed to a limit of 
$\sim10^6$ M$_{\odot}$ (Karachentsev \& Kaisin 2007), 
a few faint, compact H$\alpha$ knots have been tentatively identified
in narrowband imaging (Karachentsev \& Kaisin 2007; Paper I). 
HST stellar color-magnitude
diagrams have been used to reconstruct the
star formation history, indicating 
that the average SFR over the past Gyr is $\sim8\times10^{-4}$ \msunyr,
but with error bars consistent with no star formation (Weisz et al. 2010). 
The narrowband excess sources may therefore be background galaxies.  Follow-up spectroscopy is required to
confirm the nature of these detections.  No significant FUV emission is measurable in apertures 
spanning the extent of the galaxy, but photometry with a smaller aperture limited to the region
where the narrowband excess sources are found, yields a FUV flux that is consistent with that expected
from the H$\alpha$ emission within large uncertainties ($m_{FUV}=24.5\pm3.3$). The faint, extended diffuse emission 
observed in the NUV likely arises from older main sequence turn-off stars. 

UGCA276 belongs to a prominent association of dwarf galaxies around NGC 4214 at 2.9 Mpc (Tully et al. 2006).  Morphologically, it has been classified as a dwarf irregular.  There is a published HI flux for the system, though the detection is uncertain (de Vaucouleurs et al. 1991; Karachentsev et al. 2004).  However, there is no observable H$\alpha$ emission (Paper I).  The NUV emission in this system is also faint, extended and diffuse, so it is not likely to arise from very recent star formation.  
HST resolved stellar population imaging from ANGST also shows that UGCA276 is dominated by an old
stellar population, and has had little recent activity.  Similar to UGC5428, the reconstructed 
star formation history over the past Gyr is $\sim4\times10^{-4}$ \msunyr, but with
error bars consistent with zero (Weisz et al. 2010).
If the detection of HI is found to be robust, this would make the system extremely peculiar as UGCA276 would be the only galaxy in our sample with detectable HI gas but no evidence of recent star formation. 
Further detailed study of both objects would be useful to establish the basis of their current evolutionary status and to understand the possible impact of their group environments. 
  
In the NUV, only 10 galaxies were not detected, and are a subset of the FUV non-detections.  Half are dwarf irregulars which only have shallow exposures ($\lesssim$200 s)\footnote{HS98\_117, UGCA86, UGCA92, KKH34, LSBCD634-03}, and the other half are dwarf spheroidals.\footnote{UGCA133, IKN, DDO078, BK06N, [KK98]208}

The nearly ubiquitous presence of UV emission in local spirals and irregulars is not surprising given that they are star-forming systems, and moreover, because H$\alpha$ has already been detected in the vast majority of late-type galaxies surveyed (e.g., Meurer et al. 2006; James et al. 2008; Paper I).  This finding reinforces the idea that the intermittent cessation of activity is uncommon, as we rarely observe such systems in an ``off" mode, at least for the $B<15.5$ population probed (Lee et al. 2009a).  The result is particularly relevant for dwarf galaxies, which generally appear to have fluctuating or bursty star formation histories (e.g., Weisz et al. 2008; Tolstoy et al. 2009 and references therein), because it implies that the fluctuations do not go to zero for durations comparable to the lifetimes of UV emitting stars ($\sim$100 Myr).  For the higher luminosity galaxies in the sample ($M_B\lesssim-14$, star formation over galaxy-wide scales further appears to be virtually continuous --- the 100\% H$\alpha$ detection rate for this subset of our sample (Paper I; Lee et al. 2009a) constrains gaps in activity to be less than the lifetimes of ionizing stars ($\lesssim$ 10 Myr).

It is not clear whether this picture continues to be true for star-forming dwarfs with lower luminosities. 
In Lee et al. (2009a), it was noted that all of the late-type galaxies which were undetected in the Paper I H$\alpha$ survey had very low luminosities, with $-7.6\leq M_B \leq-13.6$.  Thus, it is possible that off-modes may be a general feature of the star formation histories of this extreme population.  However, it is not straightforward to interpret the lack of H$\alpha$ as an absence of star formation.  Such systems have lifetime-averaged SFRs on the order of $10^{-3}-10^{-4}$ \msunyr, {\it so Poisson fluctuations in the sampling of the IMF may lead to the absence of O-stars, even when continuous star formation is in fact occurring}.  The FUV should be more immune to this Poisson noise, and as we discuss below, sufficiently deep GALEX observations of H$\alpha$ non-detections show them to generally have normal levels of recent star formation activity relative to the broader population of dwarf irregular galaxies.  We may therefore speculate that extremely low-luminosity dwarfs also do not commonly undergo intermittent periods of complete inactivity which last for $\gtrsim$ 100 Myrs.  This is only a speculation because our sample is incomplete in this luminosity regime.  Obtaining H$\alpha$ and FUV observations of a large complete sample of extremely low luminosity galaxies, or better yet, of a complete sample of galaxies with low HI masses ($M_{HI}\leq10^8M_{\odot}$), is required to examine whether the rarity of UV non-detections, (or equivalently the rarity of temporarily halted star formation), persists. 

\subsection{A Re-examination of Star Formation Properties in the FUV}

In light of the finding that GALEX FUV observations
may be a more robust probe of star formation in low
density environments than typical H$\alpha$ narrowband
imaging (see introduction), it is instructive
to use our dwarf galaxy dominated sample to
re-examine commonly studied star formation 
properties which have been largely based on
H$\alpha$ measurements over the past three 
decades.  Here, we illustrate how our view of
evolutionary status of dwarf galaxies may change
by re-calculating two coarse measures
of the global star formation efficiency: the SFR per unit HI mass ($M_{HI}$),
and the SFR per unit stellar mass ($M_{\ast}$).  These
quantities provide the basis for computing timescales of 
future and past star formation, and are related to the
``gas consumption timescale'' and the ``stellar birthrate,''
respectively (e.g., Kennicutt 1983; Hunter \& Gallagher 1985; 
Kennicutt et al. 1994; van Zee et al. 1997; van Zee 2001; Skillman et al. 2003; 
Brinchmann et al. 2004; Lee et al. 2004; Karachentsev \& Kaisin 2007; James et al. 2008;
Cote et al. 2009; Lee et al. 2009a; Bothwell et al. 2009).  
In this context, it is
also particularly interesting to examine the UV properties
of late-type galaxies that were not detected in earlier H$\alpha$
imaging, and thus appeared to be devoid of
star formation.

\subsubsection{Global Star Formation Efficiencies}

In Figure~\ref{fig:gasdeplete}, the SFR is plotted against $M_{HI}$
in the left panel, and SFR/$M_{HI}$ is shown
as a function of $M_B$ on the right.  There are two
sets of points: the gray set gives the value of the 
ordinate using H$\alpha$-based SFRs, while the colored
set shows those based upon the FUV.  Only galaxies that
have both FUV and H$\alpha$ measurements are shown. The SFRs 
are computed as in Lee et al. (2009b),
where best-effort corrections have been applied for internal 
dust attenuation.\footnote{Briefly, the H$\alpha$ attenuation is measured from the Balmer
decrement, for galaxies with available integrated optical
spectroscopy.  The FUV attenuation is measured from
the total infrared-to-UV flux ratio, for galaxies with
available Spitzer IR imaging.  Otherwise, empirical scaling
relations are used to estimate the attenuation from $M_B$.  There is some current debate on whether the discrepancies between the UV and H$\alpha$ SFRs may be an artifact of uncertain attenuation corrections (e.g., Boselli et al. 2009; Meurer et al. 2010).  However, it is important to note that in the present sample of dwarf galaxies, there is a trend of decreasing H$\alpha$-to-FUV flux prior to any attenuation corrections.  Thus, the conclusions reported both here and in Lee et al. 2009b should be robust to the effects of dust.}

The HI masses are computed from
single-dish 21-cm line fluxes compiled from the literature as described in Lee et al. (2009a),
where the measurements are primarily drawn from Springob et al. (2005); the HI Parkes All Sky Survey (HIPASS) 
catalog as published in Meyer et al. (2004); and the homogenized HI compilation 
of Paturel et al. (2003) as made available through the Hyperleda database.
The lines in Figure~\ref{fig:gasdeplete} indicate values of constant SFR/$M_{HI}$, in units of the age of the universe ($t_H$=13.7 Gyr). 

The difference between the global star formation efficiencies inferred from the two SFR tracers are immediately apparent.  The H$\alpha$-based values yield significantly lower efficiencies for the late-type dwarf galaxies with $M_B\gtrsim-15$ and $M_{HI}\lesssim10^{8.5}$ relative to the more luminous and more massive spirals.  This behavior is consistent with the results of many previous authors who have reported gas depletion timescales ($\propto M_{HI}$/SFR) in excess of several Hubble times for samples of dwarf irregular galaxies (e.g., van Zee 2001; Skillman et al. 2003; Karachentsev \& Kaisin 2007; Knapen \& James 2009; Bothwell et al. 2009).  In contrast, the FUV-based values generally do not fall below $t_H^{-1}$.  
Most spiral and irregular galaxies in our Local Volume sample have SFR(UV)/HI
values on the order of $t_H^{-1}$,
with 75\% between $t_H^{-1}$ and 0.1$t_H^{-1}$.  Analogous plots for the SFR per unit stellar mass are shown in Figure~\ref{fig:ssfr}, where
the stellar masses are computed using $M_B$ and mass-to-light ratios based on the $B-V$ color, as in Bothwell et al. (2009).  Similar differences between the H$\alpha$ and UV-based values are evident.  Therefore, dwarf irregular galaxies may not be as drastically inefficient at forming stars as previously determined.

Of course, a proper analysis of star formation efficiencies requires accounting for a number of other factors that we have not considered here, for example, the molecular gas mass and the mass returned to the ISM through stellar evolutionary processes (e.g., Kennicutt et al. 1994; Bigiel et al. 2008).  
However, this exercise is not meant to be a detailed examination of efficiencies, as would be required, for example, in a study of the Schmidt Law (Schmidt 1959; Kennicutt 1998a).  
Rather, the purpose is to provide a cautionary illustration of the significant differences that can arise in our understanding of the star formation properties of galaxies in the low density regime, when the FUV is used to trace the SFR instead of H$\alpha$.  It is important to carefully re-evaluate previous conclusions regarding the evolutionary state of such galaxies that have been based on H$\alpha$ measurements.

In this context of re-examining previous H$\alpha$-based results, we also note that another similar analysis of star formation efficiencies (discussed in terms of the gas depletion and stellar mass build-up times) was recently presented by Pflamm-Altenburg \& Kroupa (2009), but under the premise that the true SFR can be recovered from H$\alpha$ measurements given their non-universal model of the stellar IMF.  Briefly, in the ``integrated galactic stellar initial mass function'' (IGIMF; Kroupa \& Weidner 2003; Weidner \& Kroupa 2005, 2006), it is assumed that: (i) all stars form in clusters; (ii) the maximum mass of a star formed in a cluster is a deterministic function of the cluster mass, and this maximum mass is lower than would be expected from random sampling of a universal IMF; and (iii) the most massive cluster that is formed in a galaxy is dependent on its SFR.  Thus, according to the IGIMF model, the stellar mass distribution of galaxies with low SFRs will be deficient in massive stars (and more deficient than would be expected from simple Poisson sampling), leading to a non-linear relationship between the SFR and the H$\alpha$ luminosity.  The results given by Figures~\ref{fig:gasdeplete} and \ref{fig:ssfr} are qualitatively consistent with those of Pflamm-Altenburg \& Kroupa (2009), in that the majority of galaxies have efficiencies greater than $t_H^{-1}$ when the UV flux is used to measure the star formation activity. 
{\it However, in our analysis, no change in the assumptions about the form of IMF was required to arrive at this conclusion.}  Instead, we make the more conservative assumption that the FUV luminosity provides a more robust measure of the recent SFR than H$\alpha$ in the low density regime because it (i) primarily originates from a more abundant population of B-stars, and thus is not as prone to stochastic effects, and (ii) is emitted directly from the stellar photospheres, and so does not suffer from possible uncertainties in photoionization of gas in low density media.  We do not yet draw a firm conclusion on the underlying cause of the H$\alpha$/UV systematic, as this requires further investigation as discussed in Lee et al. (2009b) and references therein, and in Eldridge (2010).

Finally, it should be noted that 
the optical limit of the current sample may select against 
star-forming dwarf galaxies with low UV-based efficiencies ($<t_H^{-1}$).
The $B=15.5$ limit 
corresponds to log(SFR)=-2.6, for a typical FUV-B color of
2.5 (Gil de Paz et al. 2007) at a distance of 8.5 Mpc, which 
encloses approximately half the volume of our survey.
Therefore, the apparent bound of the UV-based efficiencies
shown in Figures 4 and 5 to values greater than $t_H^{-1}$
should be tested using a
sample with a fainter optical limit, or more ideally
with a HI-selected sample complete to masses of about $10^7M_{\odot}$.

\subsubsection{UV Properties of H$\alpha$ Non-detections}
As discussed in Paper I and Lee et al. (2009a), 
late-type galaxies that are completely devoid of star formation as traced
by H$\alpha$ emission 
are rare.  In those papers, it was reported that H$\alpha$ is 
not detected in only 
22 of the 410 Local Volume galaxies that had been observed. 
Among these, 3 were classified as early-type galaxies, and
19 as dwarf irregulars.  
All of the 19 late-type H$\alpha$
non-detections have extremely low $B$ band luminosities ($M_B\gtrsim-13.6$), 
and most also have some indication of the presence of HI gas.
Therefore, we concluded that the lack of H$\alpha$ 
emission might not necessarily indicate a lack of current star formation, but
rather, may merely reflect the low probability of forming massive 
ionizing stars. 
This plausibility of this scenario can be tested with the current GALEX dataset.
On-going star formation that does not result in the 
production of ionizing stars and hence does not manifest 
itself via H$\alpha$ emission, should still
be detectable with one-orbit-depth GALEX imaging. 
This is because B-stars down to $\sim$3 M$_{\odot}$
significantly contribute to the FUV flux
and are formed at much higher frequency than
the more massive ionizing stars.

Of the 19 late-type H$\alpha$ non-detections, 12 have observations
which are at least one-orbit deep.\footnote{UGCA15, LeoT, LSBCD564-08, 
ESO349-G031, UGC7298, UGCA276, M81dwA, LGS3, KDG73, KKR03, LSBCF573-01,
BK3N}  False-color RGB images of these galaxies are shown in Figure~\ref{fig:images}.
All have measurable FUV emission
except for UGCA276, which was discussed in the preceding section.  These 11 H$\alpha$-undetected, 
UV-bright dwarf irregular galaxies are plotted in the color-magnitude
diagrams in Figure~\ref{fig:colormag2} as the circled points.  
In contrast to 
Figure~\ref{fig:colormag}, only Sc and later type spirals are now shown, 
and galaxies without H$\alpha$ observations have 
been removed to focus the comparison.  All 11 galaxies have blue UV colors 
typical of star-forming galaxies except for the
two Local Group galaxies LGS3 and LeoT, which are the lowest
luminosity galaxies in the sample.  The color of LeoT has
too large of an error for the value to be meaningful, while
LGS3's position off
of the blue sequence may be an indication that star formation has 
not occurred in the past $\sim$100 Myr.  However, with a 
lifetime averaged SFR of $\sim5\times10^{-5}$ \msunyr, even the UV emission
of these systems is susceptible to Poisson noise in the formation
of B-stars.\footnote{Following Section 4.6 of Lee et al. (2009b),
we can estimate when Poisson fluctuations
may lead to FUV non-detections, by computing
the SFR at which the number of OB stars at any
given time would be $\sim$10, as there will be times
when a galaxy with said SFR will have $10-3\sqrt{10}$
(i.e. zero) OB stars.  
Assuming a Salpeter IMF with mass range of 0.1-100 M$_{\odot}$,
95\% of the FUV light is emitted
by main sequence stars between 3.3 and 100 M$_{\odot}$.
Thus, if the IMF is integrated between 3.3 and 100 M$_{\odot}$,
the mass of stars that must be formed in order to produce
10 OB stars is 400 M$_{\odot}$.  To transform this into
a star formation rate, the mass is divided by the median lifetime
of stars that significantly contribute to the UV luminosity,
which is 100 Myr.  The limiting SFR is roughly
$4\times10^{-4}$ \msunyr.  Monte Carlo simulations
by Tremonti et al. (2007) confirm that Poisson fluctuations
do not begin to significantly affect the FUV emission until
SFRs of until at least $4\times10^{-4}$ \msunyr. \label{foot:fuvlimit}}

It happens that observations of the resolved stellar populations
of both LGS-3 (Miller et al. 2001) and Leo~T
(de Jong et al. 2008) do indicate the presence of recent star formation
(within the last few hundred Myr).  These studies of LGS-3 and Leo~T, together with 
the relatively blue UV colors of the remaining 9 
H$\alpha$  non-detections, and the generally 
clumpy morphology of their FUV emission, provide evidence that the UV emission arises
from stars that are a few hundred Myr old or younger, rather than from evolved stars 
as found in nearby elliptical galaxies (i.e., the ``UV upturn,'' O'Connell 1999 and references therein).
If the observed FUV emission
is translated into an H$\alpha$ surface brightness (assuming the standard conversion of
Kennicutt 1998b holds, and that the H$\alpha$ and UV emission extend over the same regions of the disk), the expected values
for all of the galaxies except LGS3 and LeoT are between $2\times10^{-17}$ and $1\times10^{-16}$ ergs cm$^{-2}$ s$^{-1}$ arcsec$^{-2}$
(emission measures between $\sim$ 6 and 30 pc cm$^{-6}$).  Therefore, the H$\alpha$ non-detection of these galaxies should not be due to the limiting depth of the previous narrowband observations
which probed to a 3$\sigma$ surface brightness of  $\sim1\times10^{-17}$ ergs cm$^{-2}$ s$^{-1}$ arcsec$^{-2}$.

Given that the FUV emission likely arises from young stellar
populations, the question then becomes whether the absence of HII regions 
is due to observing the galaxies during a particular time in their star formation histories
when the most massive, ionizing stars have just died (i.e.,
star formation in the galaxy has recently halted), or whether it is merely the result
of statistical fluctuations in the formation of massive stars in extremely 
low-mass systems (i.e., star formation is proceeding but  massive stars have not been produced).
Though it is difficult to unambiguously differentiate between these two 
scenarios with the current dataset alone, we note that
the H$\alpha$-undetected, UV-bright dwarf irregular galaxies all
have $M_{FUV}>-11.2$ (Figure~\ref{fig:colormag2}).  The corresponding SFR
 is 0.0018 or $10^{-2.74}$
\msunyr, which coincides with
the SFR limit below which Poisson fluctuations in the formation of
high mass ionizing stars begin to affect the H$\alpha$
output, as computed in Lee et al. (2009b).  Thus, the UV data are
at least not inconsistent with the possibility
that the non-detection of H$\alpha$ is due to
stochasticity in the sampling of the massive end of the stellar IMF 
in systems with on-going but ultra-low star formation activities.  
Again, a large, complete sample of these lowest-luminosity dwarf irregulars
would be needed to probe this issue further. 
A new generation of recently developed 
stellar population synthesis models 
that incorporate random sampling 
of a standard IMF (e.g., Eldridge et al. 2010; Fumagalli et al. 2010;
Popescu \& Hanson 2010) can be used to test whether the incidence of H$\alpha$ 
undetected, UV-bright galaxies
and the distribution of the H$\alpha$-to-FUV flux ratio,
is consistent with the expectations of stochasticity, given
a particular star formation history.

Another clue on the nature of star formation in H$\alpha$-undetected, 
UV-bright dwarf irregular galaxies is provided by examining their
UV-based star formation properties relative to those of the
overall population of late-type galaxies.  In Figures~\ref{fig:gasdeplete} and \ref{fig:ssfr}, 
where the SFR per unit HI mass and stellar mass are plotted, the
H$\alpha$ non-detections are again circled in black.
It is notable that 
most have global star formation efficiencies
that are consistent with, and not highly deviant from, the distribution 
defined by the overall population.  Aside from their extremely low luminosities
these galaxies appear to be relatively normal, at least by these global
measures of the star formation activity.  This picture is quite
different from one based upon H$\alpha$ observations alone.
In fact, the non-detection of H$\alpha$ in relatively gas-rich
dwarf galaxies can potentially be used to identify ``transition-type'' systems,
and this classification criteria was examined by Skillman
et al. (2003).  Transition-type systems are
galaxies which have observed properties
intermediate between dwarf spheroidals and dwarf irregulars, and thus,
 may  
be in a phase of transformation between the two galaxy classes (e.g.,
Mateo 1998; Grebel
et al. 2003; Cote et al. 2009).  In this context,
the absence of H$\alpha$ may be interpreted as an absence of recent
star formation, indicating that the galaxies are
possibly evolving into red and dead spheroidals, or 
experiencing a temporary interruption of star formation due 
to effects within the group environment.
However, galaxies classified as transition-types  
also have very low luminosities.  {\it The
analysis and discussion presented here clearly demonstrate that
the non-detection of H$\alpha$ in such extreme dwarfs
is a highly insufficient
criterion for identifying systems that are truly in 
a transformational state.}  An observation made
by Skillman et al. (2003), that such transition-types
appear to be quite similar to dwarf irregulars (for example,
in their spatial distribution within the group environment 
and in their gas properties),
was suggestive of this conclusion.  
An absence of detectable FUV emission in gas-rich
dwarfs would be a more compelling criterion, since
the FUV flux is robust to stochastic fluctuations
until much lower SFRs than H$\alpha$.$^{\ref{foot:fuvlimit}}$
Alternately, an H$\alpha$ non-detection accompanied by 
a red UV color ($FUV$-$NUV$$\gtrsim0.6$) in such systems
would also provide a better indication of halted 
star formation than the non-detection of H$\alpha$
alone.

\section{Summary}

The primary purpose of this paper is to provide 
GALEX NUV and FUV integrated photometry for a statistically robust
sample of star-forming galaxies within the 11 Mpc Local Volume. 
Our GALEX programs have leveraged upon previously available observations to 
mitigate the previous $\sim$4-fold  
under-representation of low-luminosity, nearby dwarfs with SFRs
lower than $\sim$0.4 \msunyr ($M_{FUV}\gtrsim-14.5$) in the GALEX
archive.  The resultant survey provides
most complete UV imaging dataset for Local Volume galaxies
currently available. 
Together with
the precursor H$\alpha$ survey and 
the Local Volume Legacy infrared observations with Spitzer 
(as described in Kennicutt et al. 2008 and Dale et al. 
2009 respectively), the GALEX observations reported here furnish a core dataset
for studying star formation and dust in the Galactic neighborhood.

We find that UV emission is present in nearly all of the late-type
galaxies observed in our sample.  This expected result supports the idea 
that the intermittent cessation of star formation is uncommon as we 
rarely observe such galaxies in an ``off" mode (Lee et al. 2009a), at least for 
the $B<15.5$ population which is the main focus of our survey. 

In the context of recent work suggesting that FUV observations
offer a more robust probe of star formation in low
density environments compared with previous H$\alpha$ narrowband
imaging, we use the dataset to re-examine two measures of 
the global star formation efficiency (SFE), the SFR per unit HI 
gas mass and the SFR per unit stellar mass.  We find that 
dwarf galaxies may not be as drastically inefficient in 
coverting gas into stars as
suggested by prior H$\alpha$ studies,
with the vast majority of galaxies in our sample
having SFEs greater than $t_H^{-1}$.  

We also examine the
properties of late-type dwarf galaxies that previously appeared 
to be devoid of star formation because they were not detected 
in earlier H$\alpha$ narrowband observations.
We find that such galaxies 
have UV SFRs that
fall below the limit where the H$\alpha$ output 
is expected to begin to suffer from
Poisson fluctuations in the formation of massive stars.
We note that the H$\alpha$-undetected, 
UV-bright systems appear to be relatively normal with respect 
to the overall population of star-forming galaxies in their
UV-based star formation efficiencies and UV colors.  
Thus, the UV data are
at least not inconsistent with the possibility
that the absence of H$\alpha$ is simply due to
stochasticity in the sampling of the massive end of the stellar IMF,
in otherwise normal star-forming systems with on-going but ultra-low star formation activities.

At a number of points in the discussion, we remark that 
obtaining FUV and H$\alpha$ observations for a deeper sample
of dwarf galaxies, or more ideally, an HI selected sample
probing masses down to $10^7$ M$_{\odot}$, is needed to
test some of our conclusions.  Such new observations
are required to: examine whether the rarity of star formation
that has been temporarily halted on $\sim$100 Myr timescales
(i.e., the rarity of UV non-detections) persists in lower
luminosity systems; test the apparent bound of SFE's to values
greater than $t_H^{-1}$; and to further probe whether the
non-detection of H$\alpha$ in UV-bright, HI-rich dwarf
galaxies can be fully explained by Poisson sampling 
of a universal IMF.

\acknowledgments
GALEX is a NASA Small Explorer, and we gratefully acknowledge NASA's support for construction, operation, and science analysis for the GALEX mission, developed in cooperation with the Centre National d'Etudes Spatiales of France and the Korean Ministry of Science and Technology. This research has made use of the NASA/IPAC Extragalactic Database (NED), which is operated by the Jet Propulsion Laboratory, California Institute of Technology, under contract with NASA, as well as the HyperLeda database (http://leda.univ-lyon1.fr).  We thank the anonymous referee for useful feedback which helped to clarify some important points in our interpretation of the data.  Special thanks are due to Chris Martin and the members of the GALEX team (http://www.galex.caltech.edu/about/team.html) for their dedicated support of the Guest Investigator Program, without which this science would not have been possible.



{\it Facilities:} \facility{Bok, CTIO:0.9m, GALEX, Spitzer, VATT}.




\clearpage
\input{table1.v2.astroph.tex}
\input{table2.v2.astroph.tex}
\input{table.lvl.astroph.tex}



\begin{figure}
\plotone{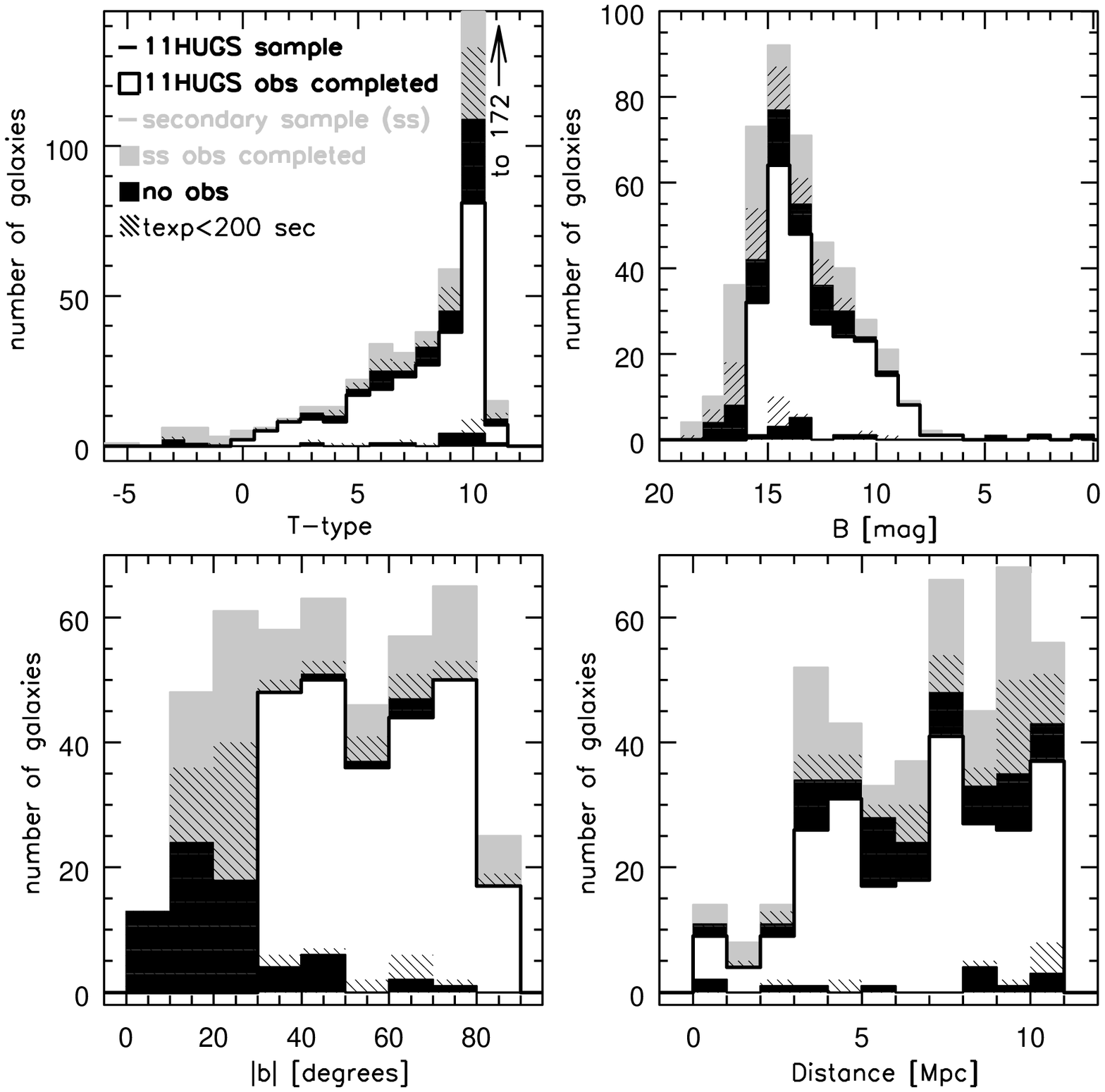}
\caption{The resultant GALEX coverage of the overall 11 Mpc sample as given in Table 1 and described in Section 2.1. Distributions in the properties used to select the sample are shown: RC3 type (top left), apparent B magnitude (top right), Galactic latitude (bottom left), and distance (bottom right).  The overall sample is given by the gray histogram, while the sub-set targeted for GALEX observations by 11HUGS ($b>30^{\circ}$, $B\leq15.5$) is distinguished in white.  The hatched areas of the histograms show the portions of each sample for which only shallow imaging exists ($t_{FUV}<200$ sec), while the black areas show that for which no FUV observations have been carried out.  For $b>30^{\circ}$ and $B\leq15.5$, there are no strong biases in the available one-orbit depth GALEX imaging in any of these properties.}
\end{figure}

\begin{figure}
\epsscale{1}
\plottwo{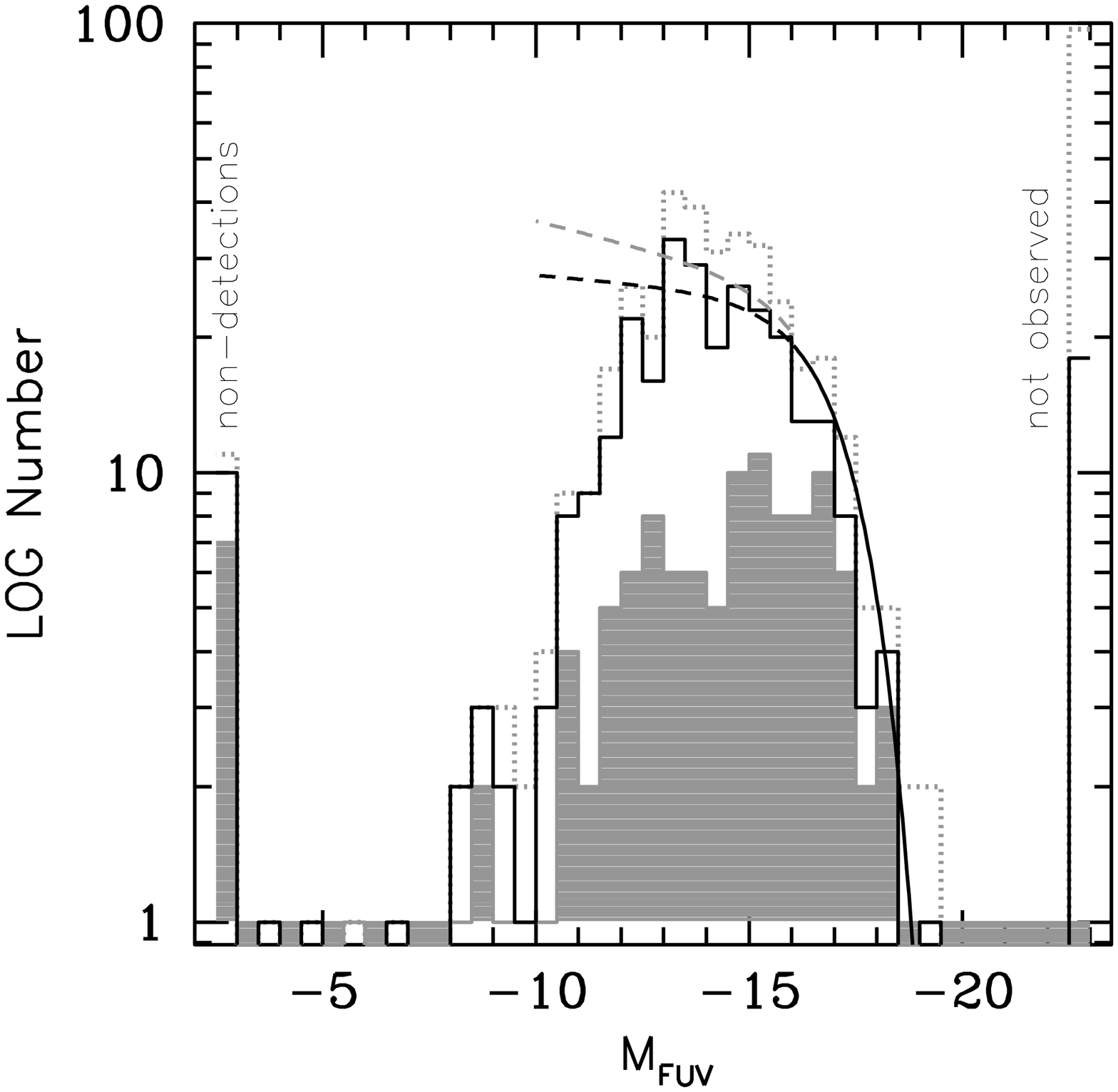}{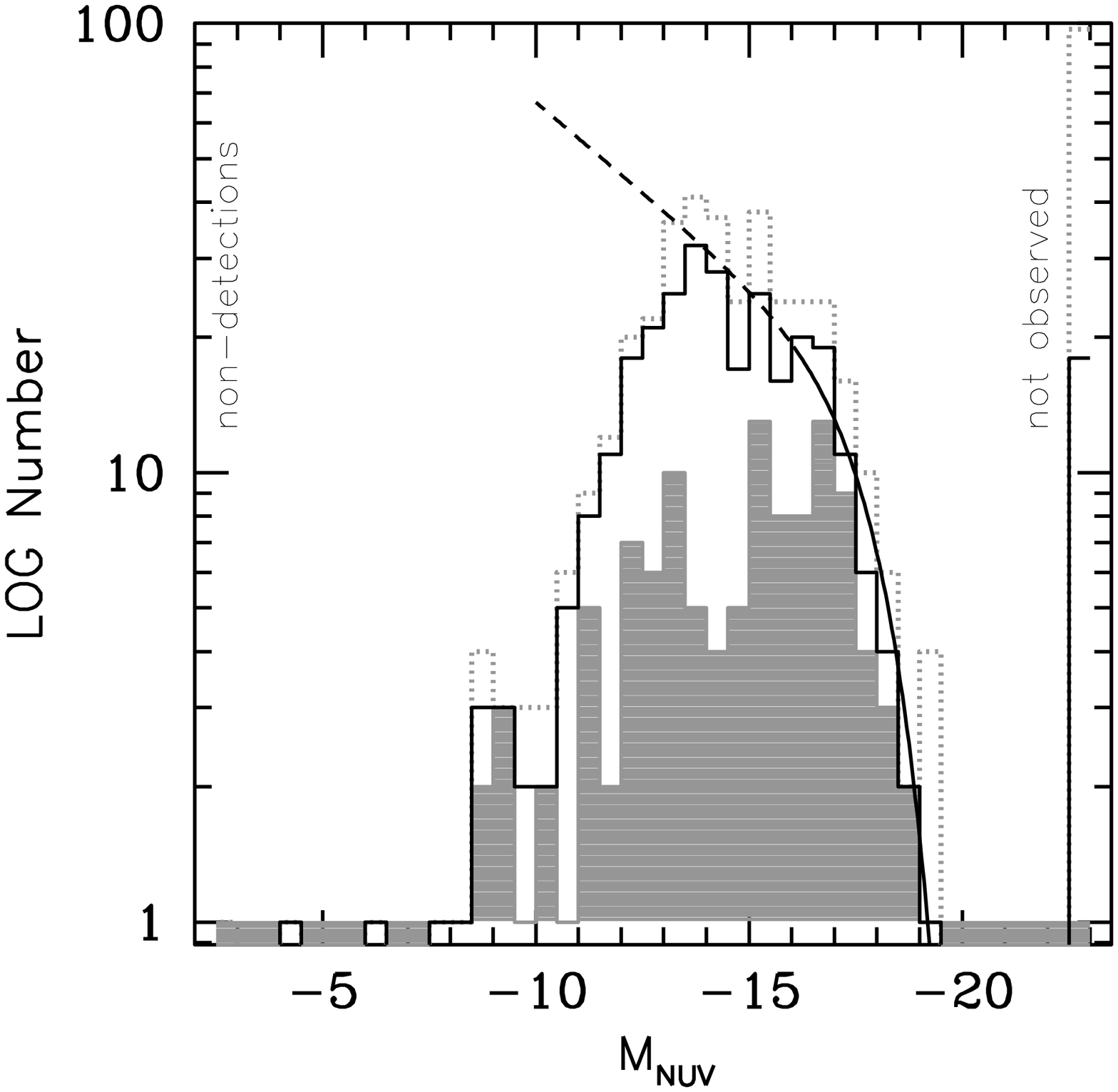}
\caption{Number distributions in NUV and FUV absolute magnitude for all galaxies with GALEX data in the overall parent (Paper I) sample (dotted gray histogram), and the subset of the sample which is most complete: spiral and irregular galaxies (T$>$0) which avoid the plane of the Milky Way $|b|> 30\degr$~(black histogram).  Corrections for foreground Milky Way extinction, but not extinction internal to the galaxies, have been applied.  The gray area shows the GALEX coverage of the $|b|>30$\degr, T$>$0 galaxies prior to observations by 11HUGS.  Non-detections are plotted at the far-left of the figure, while the number of galaxies that have not been observed by GALEX are shown at the far-right.  For comparison, we overplot Schechter fits to GALEX luminosity functions computed by Schiminovich et al. (2007) for disk-dominated galaxies in the SDSS spectroscopic sample (solid curve; the dashed portion is an extrapolation from the last data point shown).  These luminosity functions have been multiplied by twice the $|b|>30$, $d<11$ Mpc 11HUGS survey volume to provide approximate agreement with the observed number distributions (see \S\ref{section:sample_properties} for discussion).  Our GALEX program significantly improves the depth to which complete, statistical sampling of Local Volume star-forming galaxies is achieved.}\label{fig:uvlf}
\end{figure}

\begin{figure}
\epsscale{1}
\plotone{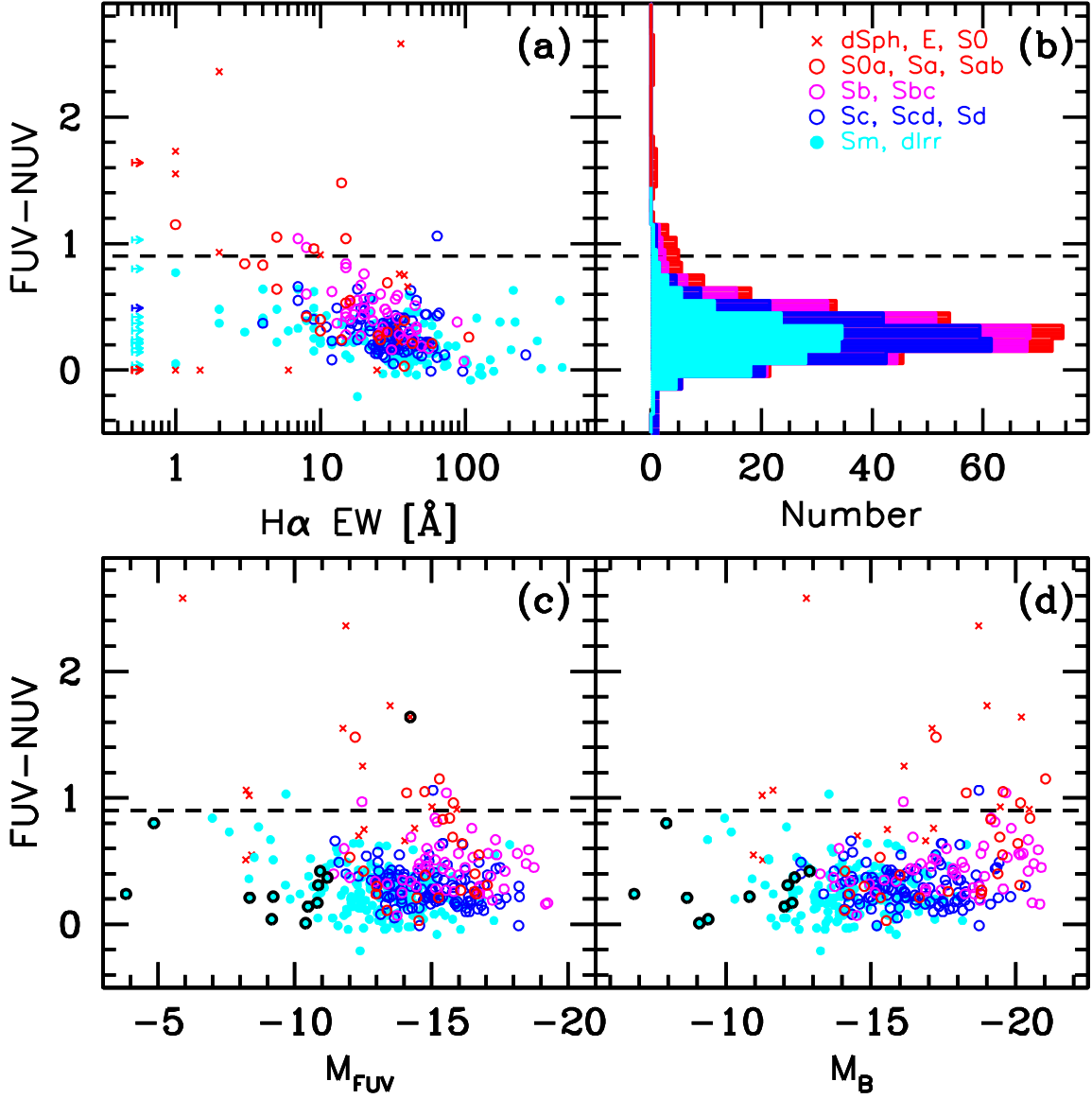}
\caption{Distributions of the integrated FUV-NUV color.  Corrections for Milky Way reddening, but not dust attenuation internal to the galaxies, have been applied. Different symbols and colors are used to distinguish between morphological type as indicated in panel (b).  Dashed lines are drawn at $FUV$-$NUV$=0.9, which Gil de Paz et al. (2007) have determined to provide good separation between early-type elliptical and lenticular ``red sequence'' galaxies, and late-type spiral and irregular ``blue sequence'' galaxies.  Galaxies which were undetected in H$\alpha$ narrowband imaging, but have on-orbit depth GALEX FUV imaging are circled in black, and largely exhibit blue UV colors characteristic of star-forming galaxies.  Blue star-forming dwarf galaxies dominate the sample by construction.}\label{fig:colormag}
\end{figure}



\begin{figure}
\epsscale{1.2}
\plottwo{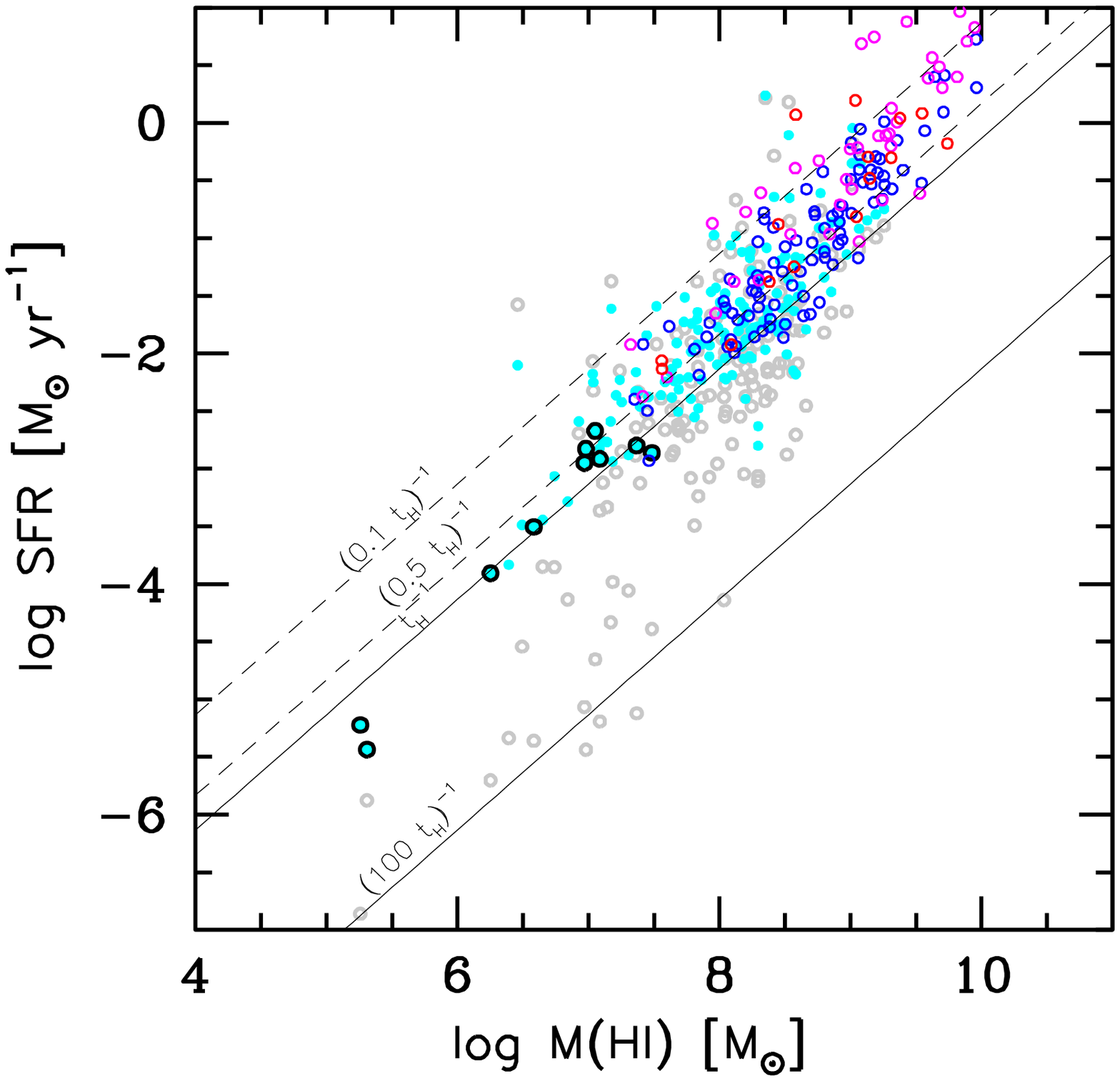}{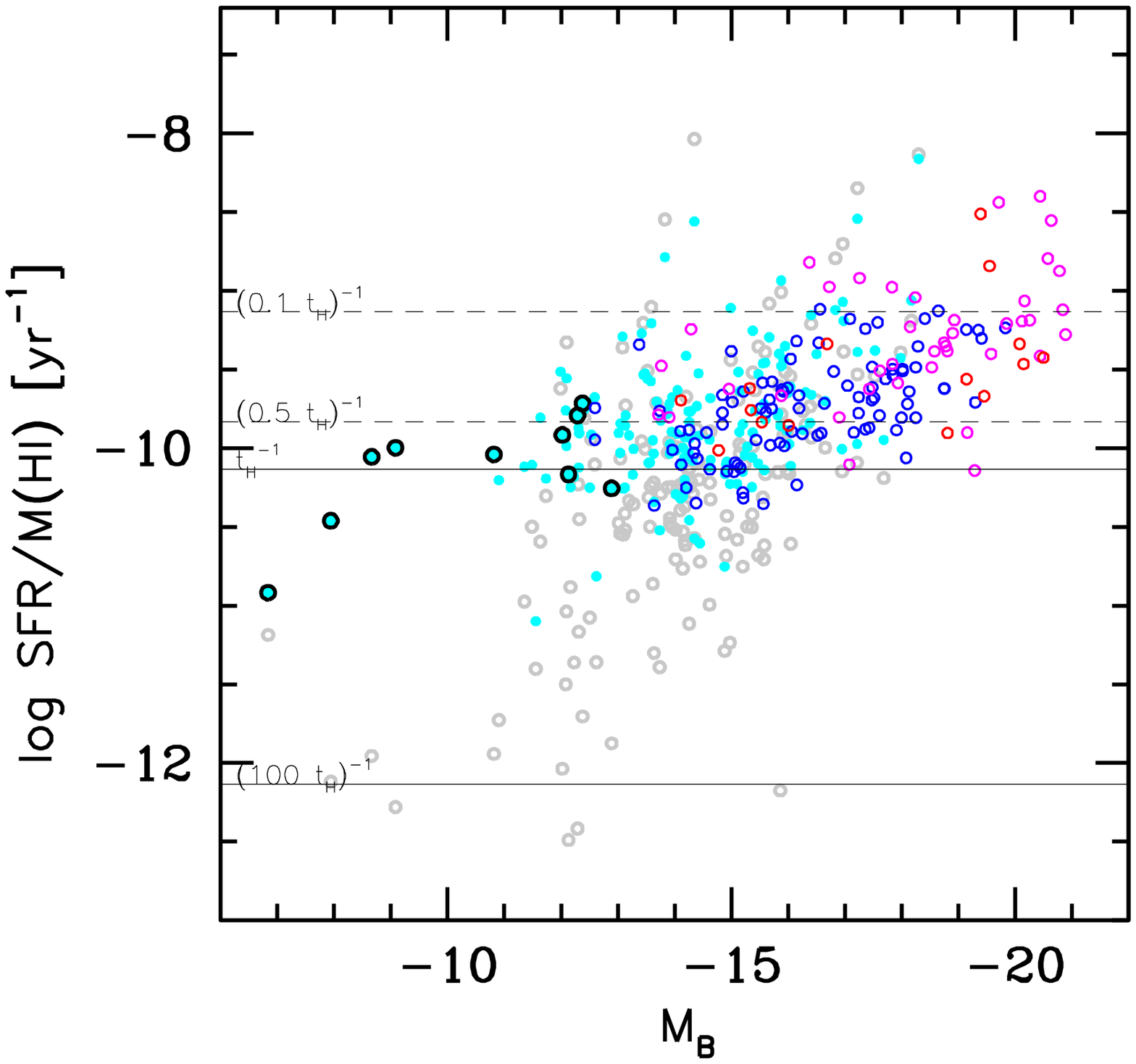}
\caption{Comparison between star formation efficiencies, as measured by the SFR per unit HI mass, when the FUV emission (colored symbols) is used as the star formation tracer, instead of H$\alpha$ (gray symbols).  Different symbols and colors are used to distinguish between morphological types as in Figure~\ref{fig:colormag}.  Only galaxies that 
have both FUV and H$\alpha$ measurements are shown.  Best effort attenuation corrections are applied before computing the SFR as in Lee et al. (2009b).}\label{fig:gasdeplete}
\end{figure}

\begin{figure}
\epsscale{1.2}
\plottwo{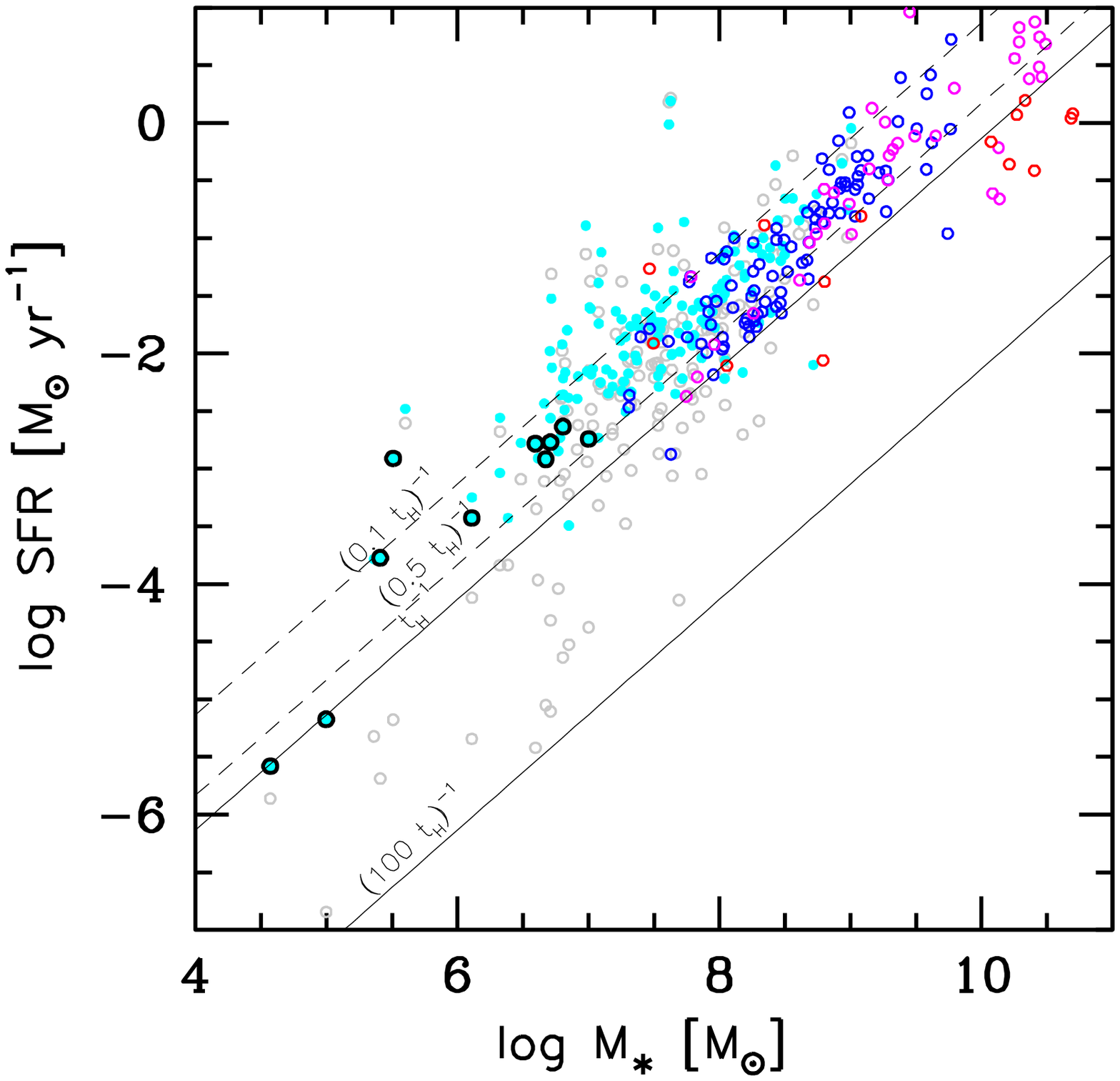}{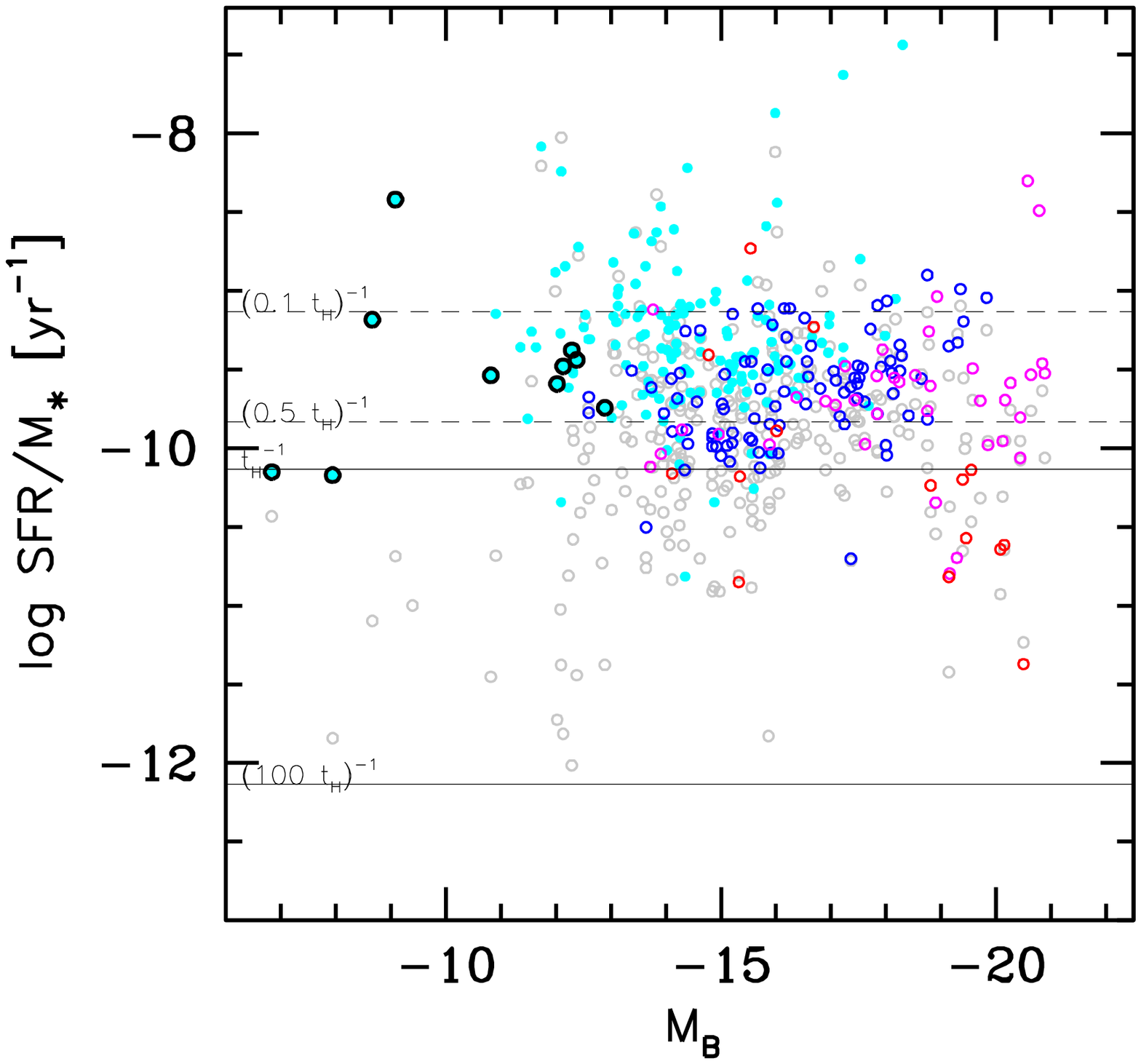}
\caption{Same as Figure~\ref{fig:gasdeplete}, but for the SFR per unit stellar mass.}\label{fig:ssfr}
\end{figure}

\clearpage
\begin{figure}

\caption{False-color images of late-type dwarf galaxies undetected in H$\alpha$ narrowband imaging which have one orbit depth GALEX data ($\sim$1500 sec).  The NUV emission appears as yellow, and the FUV emission as blue.  The green ellipses marks the RC3 $D_{25}$ ellipse.  The FUV morphologies are generally clumpy suggesting that the UV emission arises from recent star formation.  {\it (Figure 6 omitted from astro-ph posting due to size -- figure available from http://users.obs.carnegiescience.edu/jlee/papers)}  
}\label{fig:images}
\end{figure}
\clearpage
\begin{figure}

\addtocounter{figure}{-1}
\caption{False-color images of galaxies undetected in H$\alpha$ narrowband imaging, continued.}

\end{figure}

\begin{figure}
\epsscale{1}
\plotone{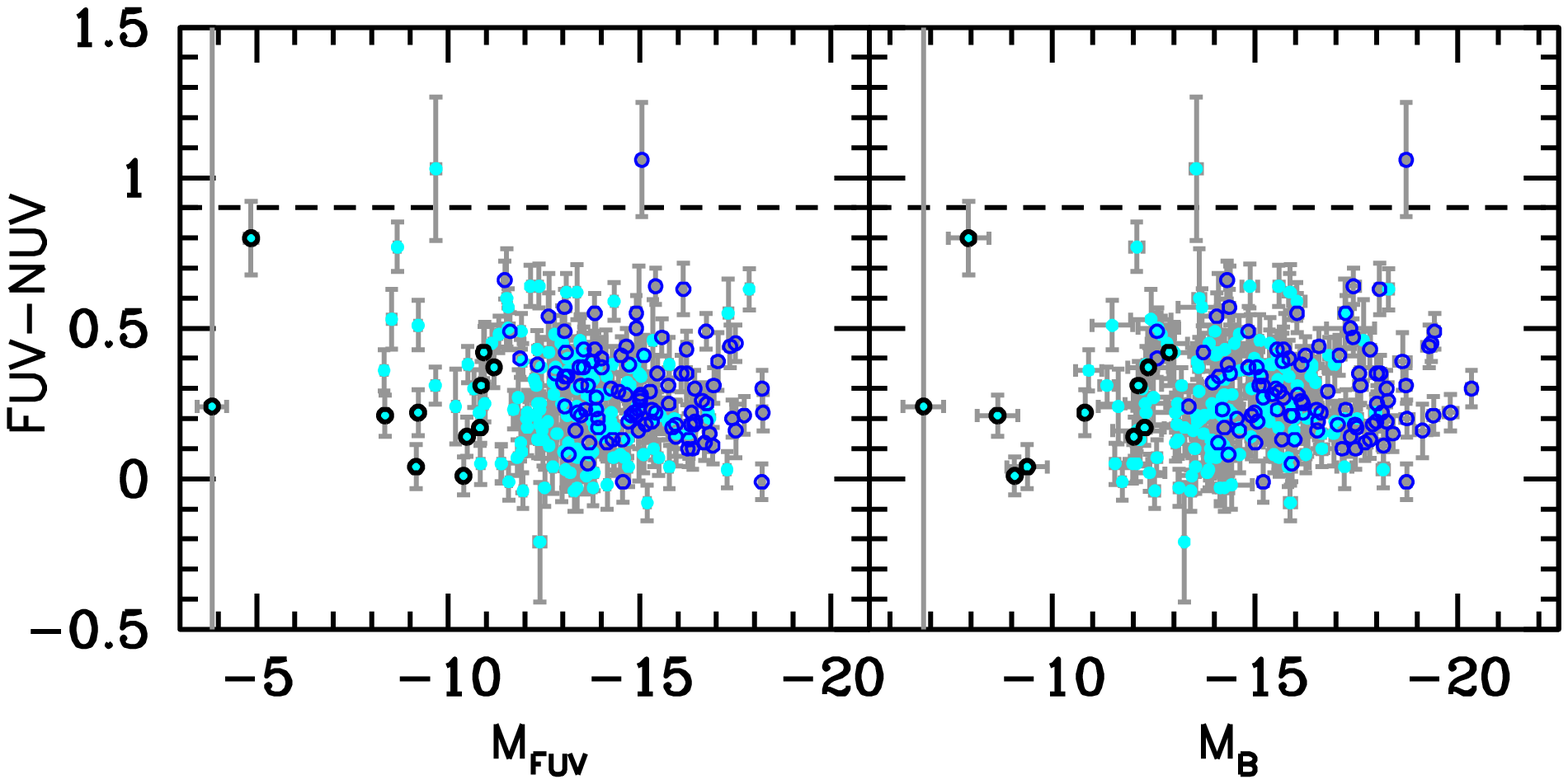}
\caption{Same as the bottom panels of \ref{fig:colormag}, but with only Sc and later-type galaxies shown.   Only galaxies that 
have both FUV and H$\alpha$ measurements are plotted.  Errors are now also shown.}\label{fig:colormag2}
\end{figure}

\clearpage




\end{document}

%% file: table1.v2.astroph.tex

\begin{deluxetable}{clcrrrclccrrc}
\tabletypesize{\scriptsize}
\rotate
\tablecolumns{12}
\tablewidth{0pc}
\tablecaption{GALEX Observations of Local Volume Galaxies}

\tablehead{

\colhead{\#}            &
\colhead{Galaxy Name}   &
\colhead{RA}            &
\colhead{DEC}           &
\colhead{$b$}           &
\colhead{D}             &
\colhead{method}        &
\colhead{$B$}           &
\colhead{T}             &
\colhead{LVL}           &
\colhead{$t_{fuv}$}      &
\colhead{$t_{nuv}$}      &
\colhead{tile name}
\\

\colhead{}              &  
\colhead{}              &
\colhead{[J2000]}       &
\colhead{[J2000]}       &
\colhead{[deg]}         &
\colhead{[Mpc]}         &
\colhead{}              &
\colhead{[mag]}         &
\colhead{}              &
\colhead{}              &
\colhead{[sec]}         &
\colhead{[sec]}         &

\\

\colhead{(1) }           &  
\colhead{(2) }           &
\colhead{(3) }           &
\colhead{(4) }           &
\colhead{(5) }           &
\colhead{(6) }           &
\colhead{(7) }           &
\colhead{(8) }           &
\colhead{(9) }           &
\colhead{(10) }          &
\colhead{(11) }          &
\colhead{(12) }          &
\colhead{(13) }          
}

\startdata
\input{table1a.astroph.dat}
\enddata
\end{deluxetable}


%% file: table1a.astroph.dat
    1 & UGC12894 & 000022.5 & 392944 &  -22.32 &  8.2 & v(flow) &  16.6 &10 &  0 &\nodata &\nodata &\nodata\\ 
     2 & WLM & 000158.1 & -152739 &  -73.63 & 0.92 & trgb & 11.03 &10 &  1 & 1440 & 1440 &NGA\_WLM\\ 
     3 & ESO409-IG015 & 000531.8 & -280553 &  -79.79 & 10.4 & v(flow) &  15.1 & 6 &  0 & 1608 & 1608 &GI1\_009016\_HPJ0005m28\\ 
     4 & ESO349-G031 & 000813.3 & -343442 &  -78.12 & 3.21 & trgb &  15.6 &10 &  0 & 1608 & 1608 &GI1\_047001\_ESO349\_G031\\ 
     5 & NGC24 & 000956.7 & -245744 &  -80.43 &  8.1 & v(flow) &  12.2 & 5 &  1 & 1579 & 1579 &NGA\_NGC0024\\ 
     6 & NGC45 & 001404.0 & -231055 &  -80.67 &  7.1 & v(flow) & 11.32 & 8 &  1 & 2220 & 2220 &GI4\_095001\_NGC0045\\ 
     7 & NGC55 & 001454.0 & -391149 &  -75.74 & 2.10 & trgb &  8.42 & 9 &  1 & 1482 & 1482 &NGA\_NGC0055\\ 
     8 & NGC59 & 001525.4 & -212642 &  -80.02 &  5.3 & sbf &  13.1 &-3 &  1 & 2393 & 2393 &GI4\_095002\_NGC0059\\ 
     9 & MCG-04-02-003 & 001911.4 & -224006 &  -81.44 &  9.8 & v(flow) &  15.6 & 9 &  0 & 1655 & 3429 &GI1\_047003\_MCG\_04\_02\_003\\ 
    10 & IC10 & 002023.1 & 591735 &   -3.34 & 0.66 & ceph &  11.8 &10 &  0 &\nodata &\nodata &\nodata\\ 
    11 & ESO473-G024 & 003122.5 & -224557 &  -83.70 &  8.0 & v(flow) & 16.04 &10 &  0 & 1600 & 1600 &GI1\_009007\_HPJ0031m22\\ 
    12 & AndIV & 004230.1 & 403433 &  -22.27 & 6.11 & trgb & 16.60 &10 &  0 & 3562 & 3562 &NGA\_M31\_MOS12\\ 
    13 & NGC224 & 004244.3 & 411609 &  -21.57 & 0.79 & ceph &  4.36 & 3 &  0 &\nodata &\nodata &\nodata\\ 
    14 & IC1574 & 004303.8 & -221449 &  -84.76 & 4.92 & trgb & 14.50 &10 &  1 & 1626 & 1627 &GI4\_095043\_IC1574\\ 
    15 & NGC247 & 004708.3 & -204538 &  -83.56 & 3.52 & trgb &  9.67 & 7 &  1 & 2507 & 2507 &NGA\_NGC0247\\ 
    16 & NGC253 & 004733.1 & -251718 &  -87.96 & 3.44 & trgb &  8.04 & 5 &  1 & 3272 & 3272 &NGA\_NGC0253\\ 
    17 & UGCA15 & 004949.2 & -210054 &  -83.88 & 3.31 & trgb &  15.4 &10 &  1 & 1513 & 1513 &GI1\_009002\_UGCA015\\ 
    18 & SMC & 005244.8 & -724943 &  -44.33 & 0.06 & ceph &  2.70 & 9 &  1 &\nodata &\nodata &\nodata\\ 
    19 & NGC300 & 005453.5 & -374100 &  -79.42 & 2.00 & ceph &  8.72 & 7 &  1 & 1628 & 1628 &NGA\_NGC0300\\ 
    20 & LGS3 & 010355.0 & 215306 &  -40.89 & 0.62 & trgb & 16.18 &99 &  0 & 1700 & 1700 &NGA\_LGS3\\ 
    21 & UGC668 & 010447.8 & 020704 &  -60.56 & 0.65 & ceph &  9.88 &10 &  1 & 1696 & 1696 &NGA\_IC1613\\ 
    22 & UGC685 & 010722.4 & 164102 &  -46.02 & 4.70 & trgb & 14.20 & 9 &  1 & 1609 & 1609 &GI1\_047004\_UGC00685\\ 
    23 & UGC695 & 010746.4 & 010349 &  -61.53 & 10.2 & v(flow) &  15.3 & 6 &  1 & 1635 & 1635 &GI1\_047005\_UGC00695\\ 
    24 & UGC891 & 012118.9 & 122443 &  -49.80 & 10.8 & v(flow) & 14.72 & 9 &  1 & 2685 & 4333 &GI1\_047006\_UGC00891\\ 
    25 & UGC1056 & 012847.3 & 164119 &  -45.26 & 10.3 & v(flow) &  14.9 &10 &  1 & 2620 & 2620 &GI4\_095003\_UGC01056\\ 
    26 & UGC1104 & 013242.5 & 181902 &  -43.47 &  7.5 & bs & 14.41 & 9 &  1 & 1168 & 2864 &GI1\_036006\_UGC01104\\ 
    27 & NGC598 & 013350.9 & 303937 &  -31.33 & 0.84 & ceph &  6.27 & 6 &  1 & 3361 & 3361 &NGA\_M33\_MOS0\\ 
    28 & NGC625 & 013504.2 & -412615 &  -73.12 & 4.07 & trgb &  11.7 & 9 &  1 & 4530 & 4530 &GI1\_047007\_NGC0625\\ 

%% file: table2.v2.astroph.tex

\begin{deluxetable}{clcrrrrlrlrlrlrl}
\tabletypesize{\scriptsize}
\rotate
\tablecolumns{16}
\tablewidth{0pc}
\tablecaption{GALEX Aperture and Asymptotic Magnitudes}

\tablehead{

\colhead{\#}            &
\colhead{Galaxy Name}   &
\colhead{E(B-V)} &
\colhead{$a$} &
\colhead{$b$} &
\colhead{PA}   &
\multicolumn{2}{c}{$FUV_{ap}$}  &
\multicolumn{2}{c}{$NUV_{ap}$}  &
\multicolumn{2}{c}{($FUV-NUV)_{ap}$} & 
\multicolumn{2}{c}{$FUV_{asym}$}  &
\multicolumn{2}{c}{$NUV_{asym}$}  
\\

\colhead{}              &  
\colhead{}              &
\colhead{}              &
\colhead{[\arcsec]}       &
\colhead{[\arcsec]}       &
\colhead{[deg]}              &
\multicolumn{2}{c}{[mag]}  &
\multicolumn{2}{c}{[mag]}  & 
\multicolumn{2}{c}{[mag]}  &
\multicolumn{2}{c}{[mag]}  &
\multicolumn{2}{c}{[mag]}  
\\

\colhead{(1) }           &  
\colhead{(2) }           &
\colhead{(3) }           &
\colhead{(4) }           &
\colhead{(5) }           &
\colhead{(6) }           &
\colhead{(7) }           &
\colhead{(8) }           &
\colhead{(9) }           &
\colhead{(10) }          &
\colhead{(11) }          &
\colhead{(12) }          &
\colhead{(13) }          &
\colhead{(14) }          &
\colhead{(15) }          &
\colhead{(16) }           
}

\startdata
\input{table2a.astroph.dat}
\enddata
\end{deluxetable}


%% file: table2a.astroph.dat
    1 & UGC12894 &    0.11 &\nodata & \nodata & \nodata &\nodata & \nodata &\nodata & \nodata &\nodata & \nodata &\nodata & \nodata &\nodata & \nodata \\ 
     2 & WLM &    0.04 &    516 &     180 &       4 &  12.48 &    0.05 &  12.35 &    0.03 &   0.13 &    0.06 &  12.49 &    0.05 &  12.38 &    0.03 \\ 
     3 & ESO409-IG015 &    0.02 &     90 &      40 &     -40 &  15.94 &    0.05 &  15.82 &    0.04 &   0.12 &    0.06 &  15.94 &    0.05 &  15.81 &    0.04 \\ 
     4 & ESO349-G031 &    0.01 &     72 &      72 &       0 &  17.05 &    0.05 &  16.91 &    0.04 &   0.14 &    0.06 &  17.04 &    0.05 &  16.88 &    0.04 \\ 
     5 & NGC24 &    0.02 &    258 &      58 &      46 &  14.04 &    0.05 &  13.77 &    0.03 &   0.27 &    0.06 &  14.01 &    0.05 &  13.75 &    0.03 \\ 
     6 & NGC45 &    0.02 &    378 &     262 &     -38 &  12.55 &    0.05 &  12.36 &    0.03 &   0.19 &    0.06 &  12.52 &    0.06 &  12.32 &    0.05 \\ 
     7 & NGC55 &    0.01 &   1452 &     251 &     -72 &  10.17 &    0.05 &   9.87 &    0.03 &   0.30 &    0.06 &  10.17 &    0.05 &   9.88 &    0.04 \\ 
     8 & NGC59 &    0.02 &    150 &      75 &     -53 &  16.06 &    0.06 &  15.31 &    0.03 &   0.75 &    0.07 &  16.08 &    0.06 &  15.31 &    0.03 \\ 
     9 & MCG-04-02-003 &    0.02 &     36 &      36 &       0 &  17.46 &    0.05 &  17.18 &    0.03 &   0.28 &    0.06 &  17.19 &    0.41 &  17.00 &    0.17 \\ 
    10 & IC10 &    1.56 &\nodata & \nodata & \nodata &\nodata & \nodata &\nodata & \nodata &\nodata & \nodata &\nodata & \nodata &\nodata & \nodata \\ 
    11 & ESO473-G024 &    0.02 &    108 &      65 &      29 &  16.75 &    0.06 &  16.71 &    0.07 &   0.04 &    0.09 &  16.77 &    0.06 &  16.75 &    0.07 \\ 
    12 & AndIV &    0.09 &     54 &      42 &     -40 &  17.65 &    0.06 &  17.15 &    0.04 &   0.50 &    0.07 &  17.37 &    0.14 &  16.48 &    0.18 \\ 
    13 & NGC224 &    0.68 &\nodata & \nodata & \nodata &\nodata & \nodata &\nodata & \nodata &\nodata & \nodata &\nodata & \nodata &\nodata & \nodata \\ 
    14 & IC1574 &    0.02 &    120 &      46 &      -5 &  16.56 &    0.05 &  16.07 &    0.03 &   0.49 &    0.06 &  16.55 &    0.05 &  16.05 &    0.03 \\ 
    15 & NGC247 &    0.02 &    954 &     307 &      -6 &  11.37 &    0.05 &  11.15 &    0.03 &   0.22 &    0.06 &  11.38 &    0.05 &  11.16 &    0.03 \\ 
    16 & NGC253 &    0.02 &   1236 &     305 &      52 &  11.16 &    0.05 &  10.63 &    0.03 &   0.53 &    0.06 &  11.16 &    0.05 &  10.64 &    0.03 \\ 
    17 & UGCA15 &    0.02 &     72 &      30 &      42 &  16.81 &    0.05 &  16.64 &    0.03 &   0.17 &    0.06 &  16.77 &    0.06 &  16.58 &    0.04 \\ 
    18 & SMC &    0.04 &\nodata & \nodata & \nodata &\nodata & \nodata &\nodata & \nodata &\nodata & \nodata &\nodata & \nodata &\nodata & \nodata \\ 
    19 & NGC300 &    0.01 &    984 &     697 &     -69 &  10.21 &    0.05 &  10.08 &    0.04 &   0.13 &    0.06 &  10.22 &    0.06 &  10.10 &    0.05 \\ 
    20 & LGS3 &    0.04 &     48 &      48 &       0 &  19.16 &    0.11 &  18.36 &    0.05 &   0.80 &    0.12 &  19.11 &    0.19 &  17.54 &    0.39 \\ 
    21 & UGC668 &    0.02 &    726 &     650 &      50 &  11.42 &    0.05 &  11.31 &    0.06 &   0.11 &    0.08 &  11.43 &    0.05 &  11.35 &    0.07 \\ 
    22 & UGC685 &    0.06 &     60 &      45 &     -60 &  16.04 &    0.05 &  15.66 &    0.03 &   0.38 &    0.06 &  16.03 &    0.05 &  15.61 &    0.03 \\ 
    23 & UGC695 &    0.03 &     78 &      64 &     -45 &  16.61 &    0.05 &  16.24 &    0.04 &   0.37 &    0.06 &  16.62 &    0.05 &  16.25 &    0.04 \\ 
    24 & UGC891 &    0.03 &    102 &      44 &      47 &  16.31 &    0.05 &  16.08 &    0.03 &   0.23 &    0.06 &  16.31 &    0.05 &  16.07 &    0.03 \\ 
    25 & UGC1056 &    0.07 &     78 &      52 &      45 &  16.63 &    0.05 &  16.25 &    0.03 &   0.38 &    0.06 &  16.64 &    0.05 &  16.25 &    0.03 \\ 
    26 & UGC1104 &    0.06 &     78 &      47 &       5 &  15.60 &    0.05 &  15.41 &    0.03 &   0.19 &    0.06 &  15.60 &    0.05 &  15.40 &    0.03 \\ 
    27 & NGC598 &    0.04 &   2070 &    1219 &      23 &   8.00 &    0.05 &   7.71 &    0.03 &   0.29 &    0.06 &   7.86 &    0.05 &   7.59 &    0.04 \\ 
    28 & NGC625 &    0.02 &    342 &     112 &     -88 &  13.72 &    0.05 &  13.34 &    0.03 &   0.38 &    0.06 &  13.76 &    0.05 &  13.39 &    0.03 \\ 

%% file: table.lvl.astroph.tex
\begin{deluxetable}{clrrrrlrl}
\tabletypesize{\scriptsize}
\tablecolumns{16}
\tablewidth{0pc}
\tablecaption{GALEX Photometry in Spitzer LVL Apertures}

\tablehead{

\colhead{\#}            &
\colhead{Galaxy Name}   &
\colhead{$a$}   &
\colhead{$b$}   &
\colhead{PA}   &
\multicolumn{2}{c}{$FUV_{LVL}$}  &
\multicolumn{2}{c}{$NUV_{LVL}$} 
\\

\colhead{}              &  
\colhead{}              &
\colhead{[\arcsec]}              &
\colhead{[\arcsec]}              &
\colhead{[deg]}              &
\multicolumn{2}{c}{[mag]}  &
\multicolumn{2}{c}{[mag]} 
\\

\colhead{(1) }           &  
\colhead{(2) }           &
\colhead{(3) }           &
\colhead{(4) }           &
\colhead{(5) }           &
\colhead{(6) }           &
\colhead{(7) }           &
\colhead{(8) }           &
\colhead{(9) }           
}

\startdata
\input{table.lvl.astroph.dat}
\enddata
\end{deluxetable}


%% file: table.lvl.astroph.dat
    2  &  WLM  &                     336 &        170 &          0 &      12.52 &       0.05 &      12.40 &       0.03\\
    5  &  NGC24  &  	             150 &        108 &        225 &      14.08 &       0.05 &      13.81 &       0.03\\
    6  &  NGC45  &  	             288 &        228 &        336 &      12.58 &       0.05 &      12.39 &       0.03\\
    7  &  NGC55  &  	            1126 &        357 &        106 &      10.16 &       0.05 &       9.86 &       0.03\\
    8  &  NGC59  &  	             128 &         90 &        302 &      16.06 &       0.06 &      15.31 &       0.03\\
   14  &  IC1574  &  	             101 &         62 &          0 &      16.56 &       0.05 &      16.08 &       0.03\\
   15  &  NGC247  &  	             738 &        290 &        352 &      11.38 &       0.05 &      11.16 &       0.03\\
   16  &  NGC253  &  	            1025 &        404 &         50 &      11.15 &       0.05 &      10.62 &       0.03\\
   17  &  UGCA15  &  	              75 &         39 &         28 &      16.78 &       0.05 &      16.60 &       0.04\\
   18  &  SMC  &  	         \nodata &    \nodata &    \nodata &    \nodata &    \nodata &    \nodata &    \nodata\\
   19  &  NGC300  &  	             754 &        564 &        114 &      10.23 &       0.05 &      10.10 &       0.03\\
   21  &  UGC668  &  	             341 &        274 &         60 &      11.70 &       0.05 &      11.67 &       0.03\\
   22  &  UGC685  &  	              90 &         74 &        122 &      16.02 &       0.06 &      15.61 &       0.04\\
   23  &  UGC695  &  	              64 &         54 &          0 &      16.64 &       0.05 &      16.27 &       0.03\\
   24  &  UGC891  &  	              97 &         59 &         42 &      16.30 &       0.05 &      16.07 &       0.03\\
   25  &  UGC1056  &  	              62 &         58 &          0 &      16.63 &       0.05 &      16.26 &       0.03\\
   26  &  UGC1104  &  	              83 &         52 &          0 &      15.60 &       0.05 &      15.41 &       0.03\\
   27  &  NGC598  &  	            2226 &       1381 &         12 &       8.00 &       0.05 &       7.72 &       0.03\\
   28  &  NGC625  &  	             250 &        128 &         90 &      13.72 &       0.05 &      13.36 &       0.03\\
   29  &  NGC628  &  	             440 &        404 &         90 &      11.68 &       0.05 &      11.40 &       0.03\\
   30  &  UGC1176  &  	             101 &         84 &         25 &      15.79 &       0.06 &      15.62 &       0.05\\
   32  &  ESO245-G005  &             179 &        126 &        318 &      13.66 &       0.05 &      13.62 &       0.03\\
   33  &  UGC1249  &  	             234 &        102 &        331 &      13.45 &       0.05 &      13.28 &       0.03\\
   34  &  NGC672  &  	             278 &        180 &         67 &      13.18 &       0.05 &      12.83 &       0.03\\
   36  &  ESO245-G007  &             144 &        120 &          0 &      16.25 &       0.10 &      15.52 &       0.04\\
   37  &  NGC784  &  	             240 &         96 &          3 &      13.91 &       0.05 &      13.51 &       0.03\\
   39  &  NGC855  &  	              95 &         86 &         68 &      15.88 &       0.11 &      15.22 &       0.04\\
   50  &  ESO115-G021  &             200 &         56 &        221 &      14.65 &       0.05 &      14.39 &       0.03\\
   55  &  ESO154-G023  &             243 &        124 &         39 &      13.78 &       0.05 &      13.52 &       0.04\\
   59  &  NGC1291  &  	             420 &        402 &          0 &      14.22 &       0.11 &      13.38 &       0.05\\
   60  &  NGC1313  &  	             448 &        347 &        338 &      10.62 &       0.07 &      10.45 &       0.04\\
   61  &  NGC1311  &  	             150 &         70 &         36 &      14.83 &       0.05 &      14.42 &       0.03\\
   64  &  UGC2716  &  	              87 &         62 &         90 &      15.91 &       0.06 &      15.43 &       0.05\\
   65  &  IC1959  &  	             126 &         57 &        330 &      14.37 &       0.05 &      14.25 &       0.03\\
   69  &  NGC1487  &  	             215 &        130 &         70 &      13.44 &       0.05 &      13.25 &       0.03\\
   72  &  NGC1510  &  	              63 &         61 &          0 &      15.09 &       0.05 &      14.87 &       0.03\\
   73  &  NGC1512  &  	             501 &        464 &         83 &      13.44 &       0.05 &      13.21 &       0.05\\
   74  &  NGC1522  &  	              76 &         50 &         37 &      15.17 &       0.05 &      15.03 &       0.03\\
   76  &  ESO483-G013  &             102 &         74 &        322 &      15.77 &       0.05 &      15.42 &       0.03\\
   83  &  ESO158-G003  &             100 &         86 &          0 &      15.47 &       0.05 &      15.21 &       0.03\\